\documentclass[12pt,aps,prb,preprint,preprintnumbers,amsmath,amssymb]{revtex4}

\usepackage{graphicx}
\usepackage{dcolumn}
\usepackage{bm}
\usepackage{epsfig}

\begin{document}


\title{Wave packet construction in two-dimensional quantum
billiards: Blueprints for the square, equilateral triangle, and
circular cases}
\author{M. A. Doncheski} 
\affiliation{Department of Physics, The Pennsylvania State
University, Mont Alto, Pennsylvania 17237}
\email{mad10@psu.edu}

\author{S. Heppelmann, R. W. Robinett, and D. C. Tussey} 
\affiliation{Department of Physics, The Pennsylvania State
University, University Park, Pennsylvania 16802}
\email{rick@phys.psu.edu}


\begin{abstract}
We present quasi-analytical and numerical calculations of Gaussian
wave packet solutions of the Schr\"odinger equation for 
two-dimensional infinite well and quantum billiard problems with 
equilateral triangle, square, and circular footprints. These cases
correspond to $N=3$, $N=4$, and $N \rightarrow \infty$ regular
polygonal billiards and infinite wells, respectively. In each case 
the energy eigenvalues and wavefunctions are given in terms of
familiar special functions. For the first two systems, we obtain
closed form expressions for the expansion coefficients for
localized Gaussian wavepackets in terms of the eigenstates of the
particular geometry. 
For the circular case, we
discuss numerical approaches. We use these results to discuss the
short-time, quasi-classical evolution in these geometries and the
structure of wave packet revivals. We also show how related
half-well problems can be easily solved in each of the three cases.
\end{abstract}

\maketitle

\section{Introduction}\label{sec:level1}
The use of wave packets to analyze the dynamics of 
quantum mechanical systems is an increasingly important aspect of
the study of the classical-quantum interface.\cite{yeazell}
Popular software packages\cite{styer_cups} can help students
visualize the evolution of quantum states (in contrast to the more
typical time-independent stationary state solutions seen in most
textbooks) by allowing students to change parameters (such as the
initial width of a wave packet). Such
visualizations are potentially important because recent
studies,\cite{singh,cataloglu_and_robinett} have suggested that
student understanding of time-development in quantum mechanics is
rather limited. 

Several authors have discussed pedagogical descriptions of various
one-dimensional (1D) quantum mechanical problems using a wave
packet approach, studying such topics as transmission and
reflection from square
barriers,\cite{goldberg_1,barrier_1,barrier_1p,edgar} linear
potential steps,\cite{barrier_2} ``bounces'' off infinite
walls,\cite{andrews,doncheski_and_robinett} and bound state wave
packets in single square wells in either position
space\cite{greenman} or momentum space,\cite{segre} in double
wells,\cite{deutchman,johnson} and in systems of relevance to
solid state physics.\cite{periodic_1,periodic_2} 
More recently, wave packet
revivals have been studied in the context of familiar 1D quantum
mechanical bound state systems such as 1D infinite wells and the
``quantum bouncer'' problem.\cite{bluhm,
robinett_revivals,styer,structures,1d_1,quantum_bouncer,
robinett_bouncer,1d_fractional,different_fractional, carpets}

An obvious extension would be the study of wave packet
propagation in two-dimensional infinite wells of various shapes
(quantum billiards). In this paper we will discuss the
construction of Gaussian-like wave packet solutions in three simple
two-dimensional (2D) geometries, namely, the square, equilateral
triangle, and circular infinite wells. We note that measurements of
conductance fluctuations in ballistic
microstructures\cite{first_experiment} have been tentatively used
to identify features in the power spectrum with particular
closed orbits in a circular and stadium billiard. More
recently, the realization of billiards\cite{second_experiment}
with ultracold atoms in arbitrarily shaped 2D boundaries confined
by optical dipole potentials has allowed the study of various
chaotic and integrable shapes such as the stadium, ellipse, and
circle.

The study of the energy eigenvalues and eigenfunctions in general
2D billiard systems can be used to probe quantum chaos, but in the
cases we discuss here, the corresponding integrable classical
motions are easily obtained and can be compared to, and contrasted
against, the short-time quantum development. 
In each case, we  make use of closed form solutions, and for the
first two systems we  show how to obtain simple expressions for the
expansion coefficients of a gaussian wavepacket in terms of the
eigenstates of the particular system; 
for the circular case we illustrate numerical results. 
We then use our results to discuss
the short-time propagation of wave packets, compare them to
classical orbits, and investigate the structure of longer-time
quantum revivals. 

The infinite square well provides an easy
introduction to the subject, and the cases of the
circular\cite{robinett_pra} and equilateral
triangle\cite{robinett_annals} quantum billiards have recently been
examined in the research literature. We will make
appropriate use of the results in
Refs.~\onlinecite{robinett_pra} and \onlinecite{robinett_annals},
but our discussions will focus on more pedagogical aspects,
including other special cases of 2D billiards obtained by
``foldings'' of these cases to make connections to the
degeneracy of the energy eigenvalues and the symmetry structure of
energy eigenstates in simple geometries. 
The comparison of these three exactly soluble 2D
infinite well problems and their quantum dynamics is a major
thrust of this paper. Note that these systems correspond to $N=4$,
$N=3$, and $N \rightarrow
\infty$ regular polygonal billiard footprints, and it is an
interesting open question how the exact quantum revivals we
find for the
$N=3$ and 4 cases change to the approximate ones for the $N
\rightarrow \infty$ (circular) case.

We begin in Sec.~II by reviewing the basic properties of Gaussian
wave packets and the general time-dependence of 1D 
bound state systems, before discussing the
1D infinite well in Sec.~III. In
Sec.~IV we consider the general time-dependence of systems
with two quantum numbers, as well as the square billiard. We
will also be able to use this geometry to examine the special case
of the
$45^{\circ}$--$45^{\circ}$--$90^{\circ}$ triangular billiard
obtained from the square well by folding along a diagonal. In
Sec.~V we briefly review the properties of the quantum solutions to
the Schr\"odinger equation in an equilateral triangle
($60^{\circ}$--$60^{\circ}$--$60^{\circ}$) infinite well (one that
is not frequently discussed in the pedagogical
literature) and discuss both the short-time quasi-classical
propagation of wave packets as well as derive the exact quantum
revivals that are present in this system. We also note that
half-well solutions for the
$30^{\circ}$--$60^{\circ}$--$90^{\circ}$ triangle are easily
obtained by folding along an axis of symmetry. Finally, we turn
our attention in Sec.~VI to the case of the circular billiard and
again discuss how information on both the classical 
periodicities and possible quantum revivals is contained in the
energy eigenvalue spectrum. In Sec.~VII we present conclusions
and some avenues for future investigation.

Although this work will focus mostly on the formalism required to
construct Gaussian wave packets in these three geometries in the
most efficient ways and will be restricted to static images of
position-space probability densities,
we have also produced
animations of the short- and long-time development of
such wave packet solutions in both the 1D and 2D infinite wells,
illustrating the short-time, quasi-classical propagation and
the structure of quantum revivals (and fractional revivals). These
animations are available at {\tt
{<}http://www.phys.psu.edu/\~{}rick/QM/qm.html{>}} and include
visualizations of the position- and momentum space probability
densities, quantum expectation values compared to classical
trajectories, and plots of the autocorrelation function for several
different initial conditions for 1D and 2D Gaussian wave packets.

\section{One-dimensional background}

\subsection{Gaussian wave packets}
For each of the geometries we consider, we will use
initial Gaussian wave packets corresponding to 
standard free particle solutions of the Schr\"odinger equation.
Although such solutions do not explicitly satisfy the boundary
conditions at the wall(s), as long as they are sufficiently far from 
any infinite wall boundary, the resulting error can be made 
exponentially small, and the solutions can be reliably expanded 
in eigenstates of the billiard system. 
The initial wave forms we will use in one dimension are given by
\begin{equation}
\langle p|G\rangle = \phi_{G}(p,0) =
\frac{\alpha^{1/2}}{\pi^{1/4}} e^{-\alpha^2(p-p_0)^2/2} \,
e^{-ipx_0/\hbar}\,,
\label{gaussian_wave_packet_momentum_space}
\end{equation}
which corresponds to 
\begin{equation}
\langle x|G\rangle = \psi_{G}(x,0) = \frac{1}{b^{1/2}\pi^{1/4}}
e^{-(x-x_0)^2/2b^2}
e^{ip_0(x-x_0)/\hbar}\,,
\label{gaussian_wave_packet}
\end{equation}
where $b \equiv \alpha \hbar$. 
These two equivalent representations of
the same Hilbert state vector, $|G\rangle$, are, of course, 
related by the Fourier transform
\begin{equation}
\psi_{G}(x,0) = \frac{1}{\sqrt{2\pi \hbar}}
\int_{-\infty}^{+\infty} \phi_{G}(p,0)\,e^{ipx/\hbar}\,dp\,.
\end{equation}
The initial expectation values for this general state are given by
\begin{equation}
\langle x\rangle_0 = x_0,
\qquad
\langle x^2 \rangle_0 = x_0^2 + \frac{b^2}{2},
\qquad
\Delta x_0 = \frac{b}{\sqrt{2}}\,,
\label{position_space_info}
\end{equation}
and
\begin{equation}
\langle p\rangle_0 = p_{0},
\qquad
\langle p^2 \rangle_0 = p_{0}^2 + \frac{\hbar^2}{2b^2},
\qquad
\Delta p_0 = \frac{\hbar }{\sqrt{2}b}
= \frac{1}{\sqrt{2} \alpha}\,.
\label{momentum_space_info}
\end{equation}
We can easily vary the initial position $x_0$, momentum $p_0$,
and width $\Delta x_0$ of the packet. 
The expectation value for the total (kinetic) energy is
given by
\begin{equation}
\langle \hat{E} \rangle = \frac{1}{2\mu } \langle p^2 \rangle
= \frac{1}{2\mu}\Bigl( p_0^2 + \frac{\hbar^2}{2b^2}\Bigr) .
\end{equation}
We will use the notation $\mu$ for the particle mass 
to avoid confusion with various quantum numbers.

If we expand such an initial state in terms of the energy
eigenstates
$u_n(x)$ of a general 1D bound
state potential, we have
\begin{subequations}
\label{expansion_in_eigenstates}
\begin{eqnarray}
\psi_{G}(x,0) &=& \sum_{n=1}^{\infty} a_n\, u_n(x),\\
\noalign{\noindent with}
a_n &=& \!\int_{-\infty}^{+\infty}\! u_n^*(x)\, \psi_{G}(x,0)\,dx .
\end{eqnarray}
\end{subequations}
The resulting $a_n$ must satisfy two constraints, namely,
\begin{subequations}
\label{constraint}
\begin{eqnarray}
\sum_{n=1}^{\infty} |a_n|^2 &=& 1 \\
\sum_{n=1}^{\infty} |a_n|^2\,E_n&=&
\langle \hat{E} \rangle = 
\frac{1}{2\mu }\Bigl( p_0^2 + \frac{\hbar^2}{2b^2}\Bigr),
\end{eqnarray}
\end{subequations}
where $E_n$ represents the quantized energies of the potential.
Equation~(\ref{constraint}) is very useful for
numerical checks on the expansion coefficients. (We note that even
if the subsequent time development of the quantum wave packet
involves ``collisions'' with an infinite wall, if the wave packet
is initially normalized to unit probability, it will remain so
through its subsequent time development. Once we normalize the
wave packet, unitarity ensures that we can ``set it and forget
it.'') The resulting time-dependence is then given by
\begin{equation}
\psi(x,t) = \sum_{n=1}^{\infty} a_n \, u_{n}(x)\, e^{-iE_nt/\hbar},
\label{time_dependence}
\end{equation}
and we know\cite{doncheski_and_robinett,robinett_revivals}
that for short times (before the initially Gaussian packet has a
chance to ``collide'' with any of the infinite wall boundaries), its
behavior will resemble that of a free-particle solution of the form
\begin{equation}
|\psi(x,t)|^2 = \frac{1}{b_t\sqrt{\pi}}
e^{-(x-x_0 - p_0t/\mu)^2/b_t^2},
\end{equation}
where
\begin{subequations} 
\begin{eqnarray}
b_t &\equiv& \alpha \hbar \sqrt{1 + (t/t_0)^2},\\
\Delta x_t &=& \Delta x_0 \sqrt{1+(t/t_0)^2},
\end{eqnarray}
\end{subequations}
with 
\begin{equation}
t_0 \equiv \frac{\mu b^2}{\hbar}
= 
\bigl(\frac{2\mu}{\hbar^2} \bigl)
\Delta x_0^2,
\label{spreading_time}
\end{equation}
which defines the time scale for wave packet spreading. 

\subsection{General time-dependence}
For the expansion of a general wave packet 
as in Eq.~(\ref{time_dependence}), we typically expand
the energy eigenvalues about the central value of the quantum 
number $n_0$ (assumed to be an integer for simplicity) 
used in the construction of the wave packet, namely,
\begin{equation}
E(n) \approx E(n_0) + E'(n_0)(n-n_0) + {1 \over 2}
E''(n_0)(n-n_0)^2 + {1 \over 6} E'''(n_0)(n-n_0)^3 +
\cdots
\label{1d_periods}
\end{equation}
We can express the time-dependence of each quantum
eigenstate as
\begin{eqnarray}
e^{-iE_nt/\hbar} & = & \exp\bigl( -i/\hbar[E(n_0)t + (n-n_0)
E'(n_0)t +
\frac{1}{2} (n-n_0)^2 E''(n_0) t + \ldots]\bigr) \nonumber
\\
& = & \exp \bigl(-i\omega_0 t - 2\pi i(n-n_0) t/T_{\rm cl}
- 2\pi i(n-n_0)^2t/T_{\rm rev} \nonumber \\
& & \qquad - 2\pi i(n-n_0)^3t/T_{\rm super} +
\ldots \bigr), \label{14}
\end{eqnarray}
in terms of which the classical period, quantum mechanical revival, 
and superrevival times are given respectively by
\begin{eqnarray}
T_{\rm cl} &=& \frac{2\pi\hbar}{|E'(n_0)|} 
\label{classical_period_definition}\\
T_{\rm rev} &=& \frac{2\pi \hbar}{|E''(n_0)|/2}\\
T_{\rm super} &=& \frac{2\pi \hbar}{|E'''(n_0)|/6}.
\label{period_definitions}
\end{eqnarray}
The first term of Eq.~(\ref{14}) is an common overall phase that
gives the same (trivial) time-dependence to each state, while the
second term describes the short-term, classical periodicity of the
bound state system because it gives $e^{-2\pi i(n-n_0)} = 1$ for
each state after one classical period, $T_{\rm cl}$. (An alternative
derivation of the form 
$T_{\rm cl} = 2\pi \hbar/|E'(n_0)|$ is given in Appendix~A using a
WKB approach.) Bound state wave packets (with the exception of the
harmonic oscillator potential for which the energy spectrum is
exactly linear in $n$) will typically spread significantly 
after a number of classical periods, entering a so-called collapsed phase,
only to reform later in the form of a quantum revival, in which the spreading
reverses itself (completely in the case of the infinite well, partially
in the case of a general potential well) and the wave packet relocalizes.
The revival time, $T_{\rm rev}$, describes the (typically longer)
time scale for that behavior. 

A standard tool for analyzing the short-term, semi-classical
periodicity, and the long-term revival structure is the
autocorrelation function defined by\cite{nauenberg} 
\begin{eqnarray}
A(t) & \equiv & \! \int_{-\infty}^{+\infty} \, \psi^*(x,t)
\psi(x,0)\,dx 
\nonumber
\\
& = & \! \int_{-\infty}^{+\infty} \, \phi^*(p,t)\, \phi(p,0)\,dp 
\label{autocorrelation_function}
\\
& = & \sum_{n=1}^{\infty} |a_n|^2 e^{iE_n t/\hbar}, \nonumber
\end{eqnarray}
which measures the degree of overlap of the initial wave function
with itself at later times (in both position- and momentum space).

\section{One-dimensional infinite well}\label{sec3}
For the 1D infinite well of width $a$, the energy eigenvalues and
eigenfunctions are given by
\begin{eqnarray}
E_n &=& \frac{p_n^2}{2\mu} = \frac{\hbar^2 \pi^2 n^2}{2\mu a^2} 
\equiv n^2 E_0 \\
u_n(x) &=& \sqrt{\frac{2}{a}} \sin\left(\frac{n\pi x}{a}\right).
\end{eqnarray}
The classical period obtained from Eq.~(\ref{classical_period_definition})
(using the identification $\mu v_n \equiv p_n = n\pi \hbar /a$) 
is given by 
\begin{equation}
\frac{dE_n}{dn} = \frac{\hbar^2 \pi^2 n}{\mu a^2} = \frac{\hbar \pi}{a}
\left(\frac{p_n}{\mu}\right) = \frac{\hbar \pi v_n}{a},
\end{equation}
which implies that
\begin{equation}
T_{\rm cl} = \frac{2\pi \hbar}{|dE/dn|} = \frac{2a}{v_n}\, ,
\end{equation}
as expected.
The revival time is given as 
\begin{equation}
T_{\rm rev} = \frac{4 \mu a^2}{\hbar \pi} 
= \frac{2\pi \hbar}{E_0}\,,
\label{1d_revival_time}
\end{equation}
and the revivals are exact in this case because $e^{-iE_nT_{\rm
rev}/\hbar} = e^{- 2\pi i n^2}= 1$ for all values of $n$. 
Because of the quadratic $n$-dependence, there are no 
higher order time scales beyond the classical period, $T_{\rm cl}$,
 and the quantum revival time, $T_{\rm rev}$.

The expansion coefficients for a general Gaussian wave packet of
the form in Eq.~(\ref{gaussian_wave_packet}) are given by
\begin{equation}
a_n = \int_{0}^{a} [u_{n}(x)]
[\psi_{G}(x,0)] dx\,,
\label{position_space_expansion_coefficient}
\end{equation}
where we can do the integrations numerically over the finite $(0,a)$
interval. We show in Fig.~1 the
results of such a numerical computation, both for the normalization, $\sum_{n=1}^{\infty} |a_n|^2$, and the average energy,
$\sum_{n=1}^{\infty} E_n\, |a_n|^2$, for zero momentum ($p_0=0$)
wave packets as we vary the initial position, $x_0$: we use here
(and elsewhere) the numerical values 
\begin{equation}
\hbar = 2\mu = a = 1\,,
\label{physical_parameters}
\end{equation}
and show results for values of $b=1/(2\sqrt{10})$ ($1/\sqrt{10}$) 
corresponding to initial spatial widths of $\Delta x_0 = 0.05$ ($0.10$) 
as shown by the
solid (dashed) curves; we used the 40 lowest-lying energy
eigenstates in the expansion. As expected, as we move the initial
wave packet near the edge (and beyond), the probability of being
inside the well decreases uniformly, and the wider wave packet
``senses'' the wall first, as shown at the bottom of Fig.~1.
Initial wave packets that are sufficiently far from either wall
can therefore be reliably ``supported'' inside the well, despite
the fact that they do not explicitly satisfy the boundary
conditions at $x=0$ and $x=a$. 

Perhaps more unexpectedly, the average energy of these wave packets 
grows rather dramatically (from the $p_0 = 0$ value of
$\langle \hat{E} \rangle = (\hbar^2/2b^2)/2\mu$) as the packet
nears the wall. This behavior is due to the expansion 
in Eq.~(\ref{expansion_in_eigenstates}) attempting to reproduce the
sharply discontinuous jump enforced by the edge of the well at
$x=a$, which requires the inclusion of many shorter wavelength and
hence higher energy/momentum component states. In fact, for a
sufficiently poorly behaved discontinuous function, the evaluation
of the kinetic energy in such a way may be
ill-defined,\cite{robinett_book} and increase without bound as
the number of component states is increased.

To avoid all such difficulties, we will henceforth assume that the
initial Gaussian is sufficiently contained within the billiard
footprint so that we make an exponentially small error by
neglecting any overlap with the region outside the 1D or 2D well,
and may ignore any discontinuity at the
wall. In practice, this condition
only requires the wave packet to be a few times
$\Delta x_0 = b/\sqrt{2}$ away from an infinite wall boundary. We
can then extend the integration region from the finite
$(0,a)$ interval to the entire 1D space, giving the (exponentially
good) approximation for the expansion coefficients
\begin{eqnarray}
\tilde{a}_{n} & \equiv & \int_{-\infty}^{+\infty} 
\left[u_{n}(x)\right]\, \left[\psi_{G}(x,0)\right]\,dx 
\nonumber \\
& = & 
\left(\frac{1}{2i}\right)
\sqrt{\frac{4b\pi}{a\sqrt{\pi}}}
[
e^{in \pi x_0/a} e^{-b^2(p_0 + n\pi \hbar/a)^2/2\hbar^2}
-
e^{-in \pi x_0/a} e^{-b^2(p_0 - n\pi \hbar /a)^2/2 \hbar^2}],
\label{approximate_expansion_coefficients}
\end{eqnarray}
because we can write 
\begin{equation}
\sin\left(\frac{n \pi x}{a}\right) = \frac{1}{2i}
(e^{in \pi x/a} - e^{-i n \pi x/a}),
\end{equation}
and we can perform Gaussian integrals such as
\begin{equation}
\int_{-\infty}^{+\infty} e^{-ax^2 -bx}\,dx 
= \sqrt{\frac{\pi}{a}} e^{b^2/4a}
\end{equation}
in closed form. This expression is very useful because it can speed
up numerical calculation involving the expansion coefficients such
as the evaluation of the autocorrelation function in 
Eq.~(\ref{autocorrelation_function}). It also accurately encodes 
the sometimes delicate interplay between the oscillatory pieces of 
the Gaussian ($e^{-ip_0 x/\hbar}$)
and the bound state ($e^{\pm i n \pi x/a}$) wavefunctions, which can
be difficult to reproduce in a purely numerical evaluation, and it
does so in a way that is valid for arbitrarily large values of
$p_0$, where the integrand would be highly oscillatory.
Finally, Eq.~(\ref{approximate_expansion_coefficients}) 
nicely illustrates how $\psi_{G}(x,0)$ and the
$u_n(x)$ must not only have an appropriate overlap in position
space, but also must have an appropriate phase relationship
between their oscillatory terms. This phase connection leads 
to the $\exp(-b^2(p_0 \pm n \pi \hbar /a)^2/2 \hbar^2)$ terms, 
which can be understood from a complementary overlap in momentum space. 
(We discuss this point in more detail, as well as the evaluation of the
$\tilde{a}_n$ in momentum space in Appendix~B.)

We can make immediate use of
Eq.~(\ref{approximate_expansion_coefficients}) by considering zero
momentum ($p_0 = 0$) wave packets; this case corresponds to placing
an object ``at rest'' inside the infinite well potential. 
 For such cases, the only natural periodicity in the
problem is the revival time in Eq.~(\ref{1d_revival_time}), because
there is no classical periodic motion. 
In this special case, the expression for 
$\tilde{a}_n$ in Eq.~(\ref{approximate_expansion_coefficients})
simplifies even further to
\begin{equation}
\tilde{a}_n = \sqrt{\frac{4 b \pi}{a \sqrt{\pi}}} 
\,
e^{-b^2 n^2 \pi^2/a^2}
\sin(\frac{n \pi x_0}{a}),
\end{equation}
which shows that for several special values of 
$x_0$ in the well,
several of the expansion coefficients will vanish for obvious
symmetry reasons. For example, for $x_0/a = 1/2$, all
of the even
$n=2,4,6,\ldots$ coefficients are zero and the only non-vanishing
terms in the expansion are the odd ones ($n = 2k+1$) which have
energies of the form
\begin{equation}
\label{30}
E_{k} = \frac{\hbar^2 \pi^2}{2\mu a^2} (2k+1)^2
= E_{0} (4k^2 + 4k +1)
= E_0 + 8 E_0 \bigl [\frac{k(k+1)}{2} \bigr]\,.
\end{equation}
The first term in Eq.~(\ref{30}) contributes only to the same
overall phase to the time-dependence of each term. The second term
is of the form
$8E_0$ times an integer and leads to revival times that are 8 times
{\it shorter} than the standard $T_{\rm rev} = 2\pi \hbar/E_0$ in
Eq.~(\ref{1d_revival_time}). Similarly, for the cases of
$x_0/a = 1/3$, 2/3, the $a_n$ with $n=3k$ vanish,
leading to special exact revivals at multiples of $T_{\rm rev}/3$
for these two initial locations. (It is also possible to construct
odd-parity wave packets that have different patterns of special
revivals at other locations in the well.)

To illustrate both the revivals and fractional revivals in special cases,
we plot in Fig.~2 the autocorrelation function, $|A(t)|^2$, 
for one revival time,
$T_{\rm rev}$, for different initial central positions, $x_0/a$,
of the zero-momentum packet 
from the center to as near one edge as we can. 
We note that in each case there is an exact revival at
$T_{\rm rev} = 2\pi
\hbar/E_0$, while for the special cases of $x_0/a = 1/2$ and $2/3$,
we find exact revivals at shorter time intervals as expected.
For $x_0/a = 0.8$, we also notice large partial (but
not exact) revivals at $0.4T_{\rm rev}$ and $0.6T_{\rm rev}$ for
similar reasons (because $\sin(4n\pi/5)$ vanishes for $n=5k$.)

We next show in Fig.~3 the effect of ``turning on'' momentum values
for an
$x_0 = a/2$ wave packet. 
For $p_0 = 0$ 
(top line), we have the special pattern of exact revivals at 
multiples of $T_{\rm rev}/8$ noted above, due to the vanishing of
the even expansion coefficients (shown in the corresponding 
$|a_n|^2$ versus $n$ plot 
in the right column). For a small, non-zero value ($p_0 = 3\pi$,
second row), only the exact revival at $T_{\rm rev}$ remains,
because the even expansion coefficients are no longer forced to
vanish. The autocorrelation function decreases somewhat more
rapidly from its initial value than in the $p_0=0$ case,
because the particle is slowly moving away from its initial
position, in addition to spreading out.

For still larger values of momentum, such as $p_{0} = 40\pi$ (third
row), we see obvious evidence for the classical periodicity and
the first appearance of fractional revivals\cite{1d_fractional}
at rational fraction multiples of
$T_{\rm rev}$. The corresponding $a_n$
now exhibit a more obvious Gaussian shape, with a spread, $\Delta
n$, which is related to the momentum variable by $p_n =n
\pi\hbar/a$, so that
\begin{equation}
\frac{\hbar}{2\Delta x_0} = \frac{\hbar}{\sqrt{2} b}
= \Delta p = \frac{\pi \hbar}{a} \Delta n,
\end{equation}
which gives $\Delta n = \ a/2\pi \Delta x_0$. (This relation can
be useful in deciding how many values of $a_{n}$ to include in a
numerical evaluation of quantities such as $A(t)$.) For even
larger momentum values (see the $p_{0} = 400\pi$ case at the
bottom, for example), the classical period becomes much shorter
than any obvious fractional revival time scale, and the shape of
the expansion coefficient distribution is unchanged (same $\Delta
n$), but simply shifted to higher values of $n$.

These results, especially the expression for the expansion
coefficients in Eq.~(\ref{approximate_expansion_coefficients}),
can be used in the square billiard we consider in
Sec.~\ref{square}, because that geometry is clearly separable into
two 1D problems.

\section{Square billiard}\label{square}
\subsection{Time-dependence for two-dimensional systems}
Systems with two quantum numbers\cite{bluhm_2d,other_2d}
with energy $E(n_1,n_2)$
offer richer possibilities for both semi-classical periodicities as
well as wave packet revivals. The extension of
Eq.~(\ref{1d_periods}) gives two possible classical periods, namely
\begin{subequations}
\label{two_classical_periods}
\begin{eqnarray}
T_{\rm cl}^{(n_1)} &=& \frac{2\pi\hbar}{|\partial
E(n_1,n_2)/\partial n_1|} \\
T_{\rm cl}^{(n_2)} &=& \frac{2\pi\hbar}{|\partial
E(n_1,n_2)/\partial n_2|}
\, ,
\end{eqnarray}
\end{subequations}
and the long-time revival structure typically
depends on three possible times given by
\begin{subequations}
\begin{eqnarray}
T_{\rm rev}^{(n_1)} &=& \frac{2\pi \hbar}{(1/2)|\partial^2
E(n_1,n_2)/\partial n_1^2|} \label{rev_11}\\
T_{\rm rev}^{(n_2)} &=& \frac{2\pi \hbar}{(1/2)|\partial^2
E(n_1,n_2)/\partial n_2^2|} \label{rev_21} \\
T_{\rm rev}^{(n_1,n_2)} &=& \frac{2\pi \hbar}{|\partial^2
E(n_1,n_2)/\partial n_1 \partial n_2|}\,,
\label{rev_22}
\end{eqnarray}
\end{subequations}
and the revival structure depends on the interplay between these three times.

Classical closed or periodic orbits, with periods given by $T_{\rm
cl}^{\rm (po)}$, are reproduced when the two classical periods are
commensurate, namely, when
\begin{equation}
p T_{\rm cl}^{(n_1)} = T_{\rm cl}^{\rm (po)} = qT_{\rm cl}^{(n_2)}
\,. \label{2d_closed_orbit_condition}
\end{equation}
The relation (\ref{2d_closed_orbit_condition}) can be interpreted
as arising from the beating of the two classical periods
against each other.\cite{bluhm_2d}

\subsection{Time-dependence for the square billiard}\label{sec4b}
For the infinite square well (with dimensions 
$L_x \times L_y = a \times a$),
the problem simplifies to two copies of a single 1D infinite well
because of the separability of the potential. For example, 
the energy eigenvalues, $E(n_x,n_y)$, 
and eigenstates, $w_{(n_x,n_y)}(x,y)$, 
are given by
\begin{subequations}
\label{2d_energies}
\begin{eqnarray}
E(n_x,n_y) &=& \frac{\hbar ^2 \pi^2 (n_x^2+n_y^2)}{2\mu a^2}, \\
\noalign{\noindent and}
w_{(n_x,n_y)}(x,y) &=& u_{(n_x)}(x) u_{(n_y)}(y),
\end{eqnarray}
\end{subequations}
where $n_x,n_y = 1,2,3,\ldots$ are the appropriate quantum numbers.
The revival times are given by Eqs.~(\ref{rev_11}) and
(\ref{rev_21}), 
 and are simply related to each other via
\begin{equation}
T_{\rm rev}^{(n_x)} = \frac{4 \mu a^2}{\hbar \pi} = T_{\rm
rev}^{(n_y)}\,,
\label{2d_revival_time}
\end{equation}
with no cross-term present. Therefore, the quantum revival
structure also is very simply
related to that of the 1D infinite well, including the
possibilities of special symmetric revivals for zero-momentum
wave packets at particular locations, such as
$(x_0,y_0) = (a/2,a/2)$ and $(a/3,2a/3)$. 
For rectangular infinite wells with incommensurate
sides ($L_x/L_y \neq p/q$), the structure of the revival 
times may be more complex.\cite{bluhm_2d,other_2d}

Compared to the simple back-and-forth motions in the 1D infinite well,
the closed or periodic orbits here are more interesting and the 
nontrivial path lengths for closed orbits, $L(p,q)$, in the square 
billiard can be readily deduced from simple geometric 
arguments,\cite{annular_billiard,robinett_po_theory} and are given
by 
\begin{equation}
L(p,q) = 2a\sqrt{p^2 + q^2}
\label{2d_billiard_orbits}
\end{equation}
where $2p$ and $2q$ count the number of ``hits'' on the horizontal
and vertical walls respectively, before returning to the starting
point in phase space. The corresponding classical periods for such
closed trajectories would be given by
\begin{equation}
T_{\rm cl}^{\rm (po)} = \frac{L(p,q)}{v_0},
\label{2d_closed_orbit_periods}
\end{equation}
where $v_0$ is the classical speed. Such orbits can be produced by 
point particles in the 2D billiard, starting from any initial
location,
$(x_0,y_0)$, inside the box, provided they are pointed
appropriately, namely in the
$\tan(\theta) =q/p$ direction. The values of $\theta = \tan^{-1}(q/p)$ 
and $T_{\rm cl}^{\rm (po)}/\tau = \sqrt{p^2+q^2}$ (where $\tau
\equiv 2a/v_0$ is the period for the simplest, back-and-forth
closed trajectory) for many of the low-lying cases are tabulated in
Table I.

The condition for periodic orbits in
Eq.~(\ref{2d_closed_orbit_condition}) can be implemented very
easily in this case, and we will examine its derivation in detail.
For such closed orbits to
occur we require that
\begin{equation}
p\bigl(\frac{2\mu a^2}{\hbar \pi n_x}\bigr) 
= p T_{\rm cl}^{(n_x)}
= q T_{\rm cl}^{(n_y)} 
= q\bigl(\frac{2\mu a^2}{\hbar \pi n_y}\bigr) , \label{39}
\end{equation}
or $n_y = n_x (q/p)$. If we substitute Eq.~(\ref{39}) into
Eq.~(\ref{2d_energies}), as well as equate the quantized total
energy, $E(n_x,n_y)$, with the classical kinetic energy, we obtain
\begin{eqnarray}
\frac{1}{2} \mu v_0^2 
\longleftrightarrow
 E(n_x,n_y) 
& = & \frac{\hbar^2 \pi^2}{2\mu a^2} ( n_x^2 + n_y^2)
\nonumber \\ & = & \frac{\hbar^2 \pi^2}{2\mu a^2} \biggl[ n_x^2 +
n_x^2
\Bigl( 
\frac{q}{p}\Bigr)^2 
\biggr] \\
& = &
\frac{\hbar^2 \pi^2}{2\mu a^2} \biggl[\frac{n_x^2
(p^2+q^2)}{p^2}\biggr],
\nonumber 
\end{eqnarray}
or
\begin{equation}
n_x = \frac{\mu a v_0}{\hbar \pi}
\frac{p}{\sqrt{p^2+q^2}}
\qquad
\mbox{and}
\qquad
n_y = \frac{\mu a v_0}{\hbar \pi}
\frac{q}{\sqrt{p^2+q^2}},
\end{equation}
so that
\begin{equation}
T_{\rm cl}^{\rm (po)} = pT_{\rm cl}^{(n_x)} = p \left(\frac{2\mu
a^2}{\hbar
\pi n_x}\right) = \frac{2a \sqrt{p^2+q^2}}{v_0}.
\label{square_box_relation}
\end{equation}
Equation~(\ref{square_box_relation}) is consistent with the purely
classical and geometrical result from
Eq.~(\ref{2d_closed_orbit_periods}).

We can visualize the appearance of these closed orbits in the time
development of the quantum solutions of the Schr\"odinger equation
by constructing 2D Gaussian wave packets of the form
\begin{equation}
\psi_{G}(x,y;t=0) = \psi_{G}(x;x_0,p_{0x},b)
\psi_{G}(y;y_0,p_{0y},b),
\label{initial_gaussian}
\end{equation}
where
\begin{equation}
\psi_{G}(x;x_0,p_{0x},b) = \frac{1}{\sqrt{b\sqrt{\pi}}} 
e^{-(x-x_0)^2/2b^2} \, e^{ip_{0x}(x-x_0)/\hbar} .
\end{equation}
A similar expression holds for $\psi_{G}(y;y_0,p_{0y},b)$. Once
again, as long as the initial location, $(x_0,y_0)$, is far away
from the edges of the potential well, such a Gaussian form can be
easily expanded in terms of the eigenstates. The
various position and momentum space expectation values are given by
analogs of Eqs.~(\ref{position_space_info}) and
(\ref{momentum_space_info}), and the expectation value of total
energy is now
\begin{equation}
\langle \hat{E} \rangle = \frac{1}{2\mu} \langle \hat{p}_x^2 
+ \hat{p}_y^2 \rangle
= \frac{1}{2\mu} \bigl[{p_{0x}}^2 + {p_{0y}}^2 +
\frac{\hbar^2}{b^2}\bigr]
\,. 
\label{gaussian_energy}
\end{equation}
The expansion coefficients also exhibit separability with
\begin{eqnarray}
\psi_{G}(x,y;t=0) & =& \sum_{n_x=1}^{\infty}\sum_{n_y=1}^{\infty}
a_{n_x} a_{n_y} u_{(n_x)}(x) u_{(n_y)}(y) e^{-iE(n_x,n_y)t/\hbar} \\
& = & 
\biggl[ \sum_{n_x=1}^{\infty} a_{n_x}
u_{(n_x)}(x)e^{-iE(n_x)t/\hbar}
\biggr]
\biggl[ \sum_{n_y=1}^{\infty} a_{n_y}
u_{(n_y)}(y)e^{-iE(n_y)t/\hbar} \biggr]
\\ \nonumber 
& = & \psi_{G}(x,t) \, \psi_{G}(y,t). \nonumber
\end{eqnarray}
with an equally simple expression for the autocorrelation
function, namely
\begin{equation}
A(t) = A_x(t)\,A_y(t).
\end{equation}

We illustrate the time-dependence of such wave packets by plotting
$|A(t)|^2 = |A_x(t)A_y(t)|^2$ versus $t$ in Fig.~4. The results
shown are for wave packets characterized by initial
positions $(x_0,y_0) = (a/2,a/2)$ and initial momenta given by
$(p_{0x}, p_{0y}) = (p_0\cos(\theta), p_{0}\sin(\theta))$ where
$p_{0} = 400\pi$ and $\theta = \tan^{-1}(p_{0y}/p_{0x})$; we
also use $\Delta x_0 = \Delta y_0 = 0.05$ and the same physical
parameters as in Eq.~(\ref{physical_parameters}). With these
values, the classical period (for the simplest back-and-forth
motion),
$T_{\rm cl} = 2a/v_0$, the spreading time, 
$t_0$ (from Eq.~(\ref{spreading_time})), 
and the revival time, $T_{\rm rev}$ (from
Eq.~(\ref{2d_revival_time})), are given by
$T_{\rm cl} = 2\mu a/p_0 = 1/(400\pi) \approx 0.8
\times 10^{-3}$, $t_0 = (2\mu/\hbar) \Delta
x_0^2 = (0.05)^2 = 2.5 \times 10^{-3}$,
and
$T_{\rm rev} = 4\mu a^2/(\hbar \pi) = 2/\pi \approx
0.64$. 
The wave packet will then exhibit a reasonable number of classical periods
before significant spreading occurs, with the revival time scale being
much larger than both (which is a typical hierarchy for the time scales
for physical revivals, such as seen in Rydberg atom
states.\cite{yeazell})

We show $|A(t)|^2$ versus $t$ in Fig.~4 for
increasing values of
$\theta$ up to $45^{\circ}$ (about which the pattern of classical
orbits and autocorrelation function data are symmetric). We
indicate the expected location of the classical closed or periodic
orbits by stars at locations of values of $T_{\rm cl}^{\rm
(po)}/\tau =
\sqrt{p^2+q^2}$ and
$\theta =
\tan^{-1}(q/p)$ corresponding to the values in Table I. (Images of 
some of the simplest classical closed orbits are also
included at the far right for the appropriate values of $\theta$.)
We note that the classical periodicities for closed orbits are
obviously reproduced in the time-development of the appropriate
quantum wave packets. We have also confirmed that varying the initial
location of the wave packet has no affect on the results shown in
Fig.~4, consistent with the classical result that the existence of
closed orbits depend only on $p,q$ through $p_{0y}/p_{0x} =
\tan(\theta) = q/p$.

\subsection{Special triangular billiards}
The energy eigenvalues and wavefunctions for a special 2D 
triangular billiard can be easily derived from those
of the infinite square well solutions in
Eq.~(\ref{2d_energies}). The standard results for the square
well, $E(n_x,n_y) = (\hbar^2 \pi^2/2\mu a^2) (n_x^2 +
n_y^2)$ and
$w_{(n_x,n_y)}(x,y) = u_{(n_x)}(x) u_{(n_y)}(y)$,
hold for any integral $n_x,n_y\geq 1$;
for $n_x=n_y$ there is a single state, while for $n_x \neq n_y$,
there is a two-fold degeneracy. Linear combinations of these 
solutions can be written in the form
\begin{subequations}
\label{triangle}
\begin{eqnarray}
w_{(n,m)}^{(-)}(x,y) & = & \frac{1}{\sqrt{2}}
\left[u_{(n)}(x)u_{(m)}(y) - u_{(m)}(x)u_{(n)}(y)\right] 
\qquad
(m \neq n) 
\label{triangle_1}
\\
w_{(n,m)}^{(+)}(x,y) & = & \frac{1}{\sqrt{2}} \left[u_{(n)}(x)u_{(m)}(y) + u_{(m)}(x)u_{(n)}(y)\right] 
\qquad
(m \neq n) 
\label{triangle_2}
\\
w_{(n,n)}^{(0)}(x,y) & = & u_{(n)}(x)u_{(n)}(y) ,
\label{triangle_3}
\end{eqnarray}
\end{subequations}
which have the same energy degeneracy, but exhibit different patterns
of nodal lines.
These alternative forms are useful because they allow one to
discuss the energy eigenvalues and 
eigenfunctions of the 45$^{\circ}$-45$^{\circ}$-90$^{\circ}$
triangle billiard\cite{other_berry,isoceles,robinett_jmp} formed by
``folding'' the square along a diagonal, because the
$w_{(n,m)}^{(-)}(x,y)$ satisfy the appropriate boundary condition
along the new hypotenuse. (We can easily see from
Eq.~(\ref{triangle_1}) that $w_{(n,m)}^{(-)}(x,y=x) = 0$ by construction.) 
Additional foldings along diagonals are also possible,\cite{robinett_jmp} 
and the energy spectrum can be used to analyze these cases as well. 
The allowed eigenvalues for this case are still given by 
$E(n_x,n_y) = (\hbar^2 \pi^2/2\mu a^2) (n_x^2 + n_y^2)$, which
corresponds to
\begin{equation}
E(n,m) = \frac{\hbar^2 \pi^2}{2\mu a^2} (n^2 + m^2),
\end{equation}
but now with only a single $(n_x,n_y) \rightarrow (n,m)$ state
allowed, with corresponding wavefunctions given by
Eq.~(\ref{triangle_1}), but multiplied by $\sqrt{2}$ to account for
the different normalization needed in the smaller area billiard.

The revival time in the $45^{\circ}$-$45^{\circ}$-$90^{\circ}$
triangular billiard is still given by
Eq.~(\ref{2d_revival_time}) and localized wave packets can also be
constructed using the appropriately normalized analogs of the
wavefunctions in Eq.~(\ref{triangle}), again provided they
are kept away from any of the infinite wall boundaries. 

The structure of the classical closed or periodic orbits in this case
is the same as for the square billiard, because all of the standard
$(p,q)$ orbits in the square well are simply reflected off the
new diagonal wall (along the hypotenuse), giving rise to the same
allowed orbits as in Eq.~(\ref{2d_billiard_orbits}) and Table~I. 
The only new feature in the semi-classical propagation
of such wave packet solutions in this folded
geometry\cite{robinett_jmp} is the existence of a special isolated
closed orbit at $135^{\circ}$ (one that bisects the $90^{\circ}$
right angle, bouncing normally off the hypotenuse, and returning
to the corner), which has path lengths that are multiples of 
$(\sqrt{1^2 + 1^2})a/2 = \sqrt{2}a/2$, namely half that of the standard
$(p,q) = (1,1)$ features. When we construct wave packets using
parameters appropriate to this geometry, we see twice as many features
in the $A(t)$ plot for this case because of this special classical
closed orbit, but we otherwise reproduce the results shown in
Fig.~4.

\section{Equilateral triangle billiards}
It is perhaps under appreciated that 
the energy eigenvalues and position space wave functions for a
particle of mass $\mu$ in an equilateral triangular
($60^{\circ}$-$60^{\circ}$-$60^{\circ}$) infinite well (or
billiard) of side 
$a$ are available in closed form. They have been derived 
for a variety of different contexts by at least four
groups\cite{canadian,berry,math_methods,blinder_version} using
complementary methods of derivation. 

For definiteness in what follows, 
we will assume such a triangular billiard with vertices 
located at 
$(0,0)$, $(a/2,\sqrt{3}a/2)$, and $(-a/2,\sqrt{3}a/2)$. 
The resulting energy spectrum is given by
\begin{equation}
E(m,n) = \frac{\hbar^2}{2\mu a^2} \bigl(\frac{4 \pi}{3}\bigr)^2
( m^2 + n^2 - mn)
\label{energy_eigenvalues}
\end{equation}
for integral values of $m$ and $n$, with the restriction that $m
\geq 2n$. (In what follows, we will use the notation of
Ref.~\onlinecite{berry}, except for a minor relabeling of the
energies and wavefunctions: we will continue to use $\mu$ for the
particle mass to avoid confusion with various quantum numbers.) 
For $m > 2n$, there are two degenerate states with
different symmetry properties\cite{berry} that can be written in
the forms
\begin{eqnarray}
w_{(m,n)}^{(-)}(x,y) & = &
\sqrt{\frac{16}{a^2 3\sqrt{3}}}
\left[
\sin\left(\frac{2\pi (2m-n)x}{3a}\right) 
\sin\left(\frac{2\pi ny}{\sqrt{3}a}\right) 
\right. \nonumber \\ 
& & 
\qquad
- 
\sin\left(\frac{2\pi (2n-m)x}{3a}\right) 
\sin\left(\frac{2\pi my}{\sqrt{3}a}\right) 
\label{odd_wavefunctions}
\\
& &
\qquad
\left.
- 
\sin\left(\frac{2\pi (m+n) x}{3a}\right)
\sin\left(\frac{2\pi (m-n) y}{\sqrt{3}a}\right)
\right], \nonumber 
\end{eqnarray}
and
\begin{eqnarray}
w_{(m,n)}^{(+)}(x,y) & = &
\sqrt{\frac{16}{a^2 3\sqrt{3}}}
\left[
\cos\left(\frac{2\pi (2m-n)x}{3a}\right) 
\sin\left(\frac{2\pi ny}{\sqrt{3}a}\right) 
\right. \nonumber \\ 
& & 
\qquad 
- 
\cos\left(\frac{2\pi (2n-m)x}{3a}\right) 
\sin\left(\frac{2\pi my}{\sqrt{3}a}\right) 
\label{even_wavefunctions} \\
& &
\qquad 
\left.
+
\cos\left(\frac{2\pi (m+n) x}{3a}\right)
\sin\left(\frac{2\pi (m-n) y}{\sqrt{3}a}\right)
\right].
\nonumber 
\end{eqnarray}
We can confirm by direct differentiation that they satisfy the
2D Schr\"odinger equation, with the energy eigenvalues in 
Eq.~(\ref{energy_eigenvalues}), as well as the appropriate boundary 
conditions. 
We have here also included the correct 
normalizations, because we will eventually expand Gaussian wave
packets in these eigenstates. 

For the special case of $m=2n$, there is a single non-degenerate
state for each value of $n$ given by 
\begin{equation}
w_{(2n,n)}^{(0)}(x,y) = 
\sqrt{\frac{8}{a^2 3\sqrt{3}}}
\left[
2\cos\left(\frac{2\pi nx}{a}\right) \sin\left(\frac{2\pi n y}{\sqrt{3}a}\right)
- \sin\left(\frac{4 \pi ny}{\sqrt{3}a}\right)
\right] \,.
\label{special_wavefunctions}
\end{equation}
Clearly these states satisfy
\begin{subequations}
\begin{eqnarray}
w_{(m,n)}^{(\pm)}(-x,y) &=& 
\pm w_{(m,n)}^{(\pm)}(x,y)\\
w_{(m,n)}^{(0)}(-x,y) &=& 
+ w_{(m,n)}^{(0)}(x,y).
\end{eqnarray}
\end{subequations}
and the $w_{(m=2n,n)}^{(\pm)}(x,y)$ states also satisfy
\begin{subequations}
\begin{eqnarray}
w_{(m=2n,n)}^{(+)}(x,y) & =& \sqrt{2} w_{(2n,n)}^{(0)}(x,y) \\
w_{(m=2n,n)}^{(-)}(x,y) & =& 0 \,.
\end{eqnarray}
\end{subequations}
The pattern of energy level degeneracies and wavefunction symmetries
is thus very similar to that for the square
billiard, especially when the solutions for that problem are written
in the form of Eq.~(\ref{triangle}). 

We note that the corresponding momentum space wavefunctions,
$f_{(m,n)}^{(\pm)}(p_x,p_y)$, can also be obtained in closed form, 
because the required 2D Fourier transform over 
the triangular region can be done using standard integrals involving
trigonometric functions and powers, making this problem very useful
for still another reason.

We turn now to the time-dependence of wave packets in this
geometry and consider the long-term, revival time scales in this
two quantum number system:
\begin{subequations}
\begin{eqnarray}
T_{\rm rev}^{(m)} &=& \frac{2\pi \hbar}{|\partial^2 E/\partial
m^2|/2}\\
T_{\rm rev}^{(n)} &=& \frac{2\pi \hbar}{|\partial^2
E/\partial n^2|/2} \\
T_{\rm rev}^{(m,n)} &=& \frac{2\pi \hbar}{|\partial^2 E/\partial m
\partial n|},
\end{eqnarray}
\end{subequations}
which gives
\begin{equation}
T_{\rm rev}^{(m)} 
= 
T_{\rm rev}^{(n)} 
=
T_{\rm rev}^{(m,n)} 
\equiv
T_{\rm rev}
= \frac{9 \mu a^2}{4\hbar \pi}.
\label{equilateral_triangle_revival_time}
\end{equation}
Exact revivals are present in
this system (just as for the square well), with a single revival
time guaranteed for any possible initial wave packets.
Thus, both the $N=3$ and $N=4$ polygonal billiards (the
equilateral triangle and square) exhibit similar and simple energy
eigenvalues and exact quantum revivals. We note that special,
shorter-time scale revivals are also possible in the equilateral
triangle case for zero-momentum wave packets initially centered at
symmetric locations within the triangular
billiard,\cite{robinett_annals} such as at the geometrical center
and half-way down a bisector.

The short-time classical periodicity of quantum wave packets in
this geometry can also be determined from calculations of
\begin{subequations}
\label{new_shorty}
\begin{eqnarray}
T_{\rm cl}^{(m)} &=&
\frac{2\pi \hbar}{|\partial E/\partial m|}
=
\left[\frac{9\mu a^2}{4\hbar \pi} \right]\frac{1}{(2m-n)}\\
T_{\rm cl}^{(n)} &=& 
\frac{2\pi \hbar}{|\partial E/\partial n|}
=
\left[\frac{9\mu a^2}{4\hbar \pi} \right] \frac{1}{(2n-m)}
\,. 
\end{eqnarray}
\end{subequations}
The condition leading to closed orbits can then be written as
\begin{subequations}
\begin{eqnarray}
\frac{(2m-n)}{(2n-m)} &=& \frac{T_{\rm cl}^{(n)}}{T_{\rm cl}^{(m)}}
= \frac{p}{q}\\
\noalign{\noindent or}
n &=& m \left(\frac{2p+q}{2q+p} \right).
\end{eqnarray}
\end{subequations}
If we substitute this condition into the energy spectrum in 
Eq.~(\ref{energy_eigenvalues}) and equate the quantum
energies with 
the classical kinetic energy, $\mu v_0^2/2$, we are led to the
association of $\mu v_0^2/2$ with 
$E(m,n)$ and 
\begin{equation}
E(m,n) = \left(\frac{16 \pi^2}{9}\right) \left(\frac{\hbar^2}{2\mu
a^2}\right)
\left[ m^2 + m^2\left(\frac{2p+q}{2q+p}\right)^2 
- m^2 \left(\frac{2p+q}{2q+p}\right) \right],
\end{equation}
or
\begin{equation}
\bigl(\frac{2\mu v_0a}{4\pi \hbar}\bigr)^2
= m^2 \bigl[\frac{3(p^2+pq + q^2)}{(2q+p)^2} \bigr]\,. 
\end{equation}
This identification
implies that 
\begin{subequations}
\begin{eqnarray}
m &=& \left(\frac{2\mu v_0a}{4\pi \hbar} \right)
\frac{(2q+p)}{\sqrt{3}\sqrt{p^2 + pq + q^2}}\\
n &=& \left(\frac{2\mu v_0a}{4\pi \hbar} \right)
\frac{(2p+q)}{\sqrt{3}\sqrt{p^2 + pq + q^2}}
\,.
\end{eqnarray}
\end{subequations}
The period for classical, closed/periodic orbits is then given by 
\begin{equation}
T_{\rm cl}^{\rm (po)} = pT_{\rm cl}^{(m)}
= \frac{a \sqrt{3}\sqrt{p^2 + pq + q^2}}{v_0}
=
\frac{L(p,q)}{v_0},
\label{closed_orbit_period}
\end{equation}
where 
\begin{equation}
L(p,q) = d(p,q) = a\sqrt{3}\sqrt{p^2 + pq + q^2} 
\end{equation}
are the corresponding path lengths for periodic orbits. 
The possible classical closed or periodic orbits can be
derived\cite{robinett_annals} from a geometrical construction 
(involving tiling of the 2D plane, just as for the square
case\cite{annular_billiard,robinett_po_theory}) giving the
same result. We note that the special cases of $2m=n$ and $2n=m$
correspond to $q=0$ and $p=0$ respectively, both of which give
$L(p,q) = \sqrt{3}a$. In these cases, one of the classical periods 
from Eq.~({\ref{new_shorty}) 
formally goes to infinity, which can be understood classically from the
corresponding path length, which corresponds to periodic back and
forth motion from one corner, along a bisector, to the opposite
side, but with no repetition in the complementary direction. (See
Ref.~\onlinecite{berry} for a derivation of the quantized
energies from which this effect also can be inferred.)

Because of the (relatively) simple form of the allowed wavefunctions
in Eqs.~(\ref{odd_wavefunctions}) and (\ref{even_wavefunctions}), 
we can evaluate the expansion coefficients for any 2D Gaussian
wave packet of the form in Eq.~(\ref{initial_gaussian}) by
extending the region of integration from the (finite) triangular
region to the entire 2D space, as long as the initial wave packet
is far away from any of the walls.\cite{robinett_annals} 
The required Gaussian-type integrals have the forms
\begin{eqnarray}
\int_{-\infty}^{+\infty} e^{ip_0(x-x_0)/\hbar}\, 
e^{-(x-x_0)^2/2b^2}\,\cos\bigl(\frac{Cx}{a}\bigr)\,dx
& = &
\frac{b\sqrt{2\pi}}{2}
\left[
e^{iCx_0/a} e^{-b^2(C/a + p_0/\hbar)^2/2} \right.
 \\
& & 
\quad
\left. 
+
e^{-iCx_0/a} e^{-b^2(-C/a + p_0/\hbar)^2/2}
\right],
\nonumber
 \end{eqnarray}
and
\begin{eqnarray}
\int_{-\infty}^{+\infty} e^{ip_0(x-x_0)/\hbar}\, 
e^{-(x-x_0)^2/2b^2}\,\sin\bigl(\frac{Cx}{a}\bigr)\,dx
& = &
\frac{b\sqrt{2\pi}}{2i}
\left[
e^{iCx_0/a} e^{-b^2(C/a + p_0/\hbar)^2/2} \right.
 \\
& & 
\quad
\left. 
-
e^{-iCx_0/a} e^{-b^2(-C/a + p_0/\hbar)^2/2}
\right],
\nonumber
\end{eqnarray}
with similar expressions for the $y$-integrations. The resulting
closed form expressions for the expansion coefficients can be used
in calculations of the autocorrelation function to illustrate
the long-time revival structure of the wave packets, as well as the
short-time, semi-classical propagation giving rise to closed orbits
of the form in Eq.~(\ref{closed_orbit_period}); 
the analogs of Figs.~2 and 4
for the equilateral triangle billiard have been presented in 
Ref.~\onlinecite{robinett_annals}.

A folding of the equilateral
($60^{\circ}$-$60^{\circ}$-$60^{\circ}$) triangle along a bisector
yields another special triangular geometry, namely a
$30^{\circ}$-$60^{\circ}$-$90^{\circ}$ right triangle. The energy
eigenvalues and eigenfunctions for this case can also be trivially
obtained from those of Eq.~(\ref{odd_wavefunctions}) as they
satisfy the new boundary condition along the fold. Such a system
will have the same energy eigenvalues as in
Eq.~(\ref{energy_eigenvalues}) (but with only one possible
$(m,n)$ combination with $m > 2n$) and the same
common revival time, $T_{\rm rev}$, as in 
Eq.~(\ref{equilateral_triangle_revival_time}). Wave packet solutions
can also be constructed (remembering to include an additional factor of
$\sqrt{2}$ to account for the normalization difference) from the
$w_{(m,n)}^{(-)}(x,y)$ in Eq.~(\ref{odd_wavefunctions}).

\section{Circular billiards}

\subsection{Eigenstate solutions and connections
to classical closed orbits}\label{sec:closed}

We finally turn our attention to the $N \rightarrow \infty$ limit
of the $N$-sided regular polygonal billiard, namely the circular
infinite well. (We note that the problem of scattering from an
infinite cylindrical barrier in two dimensions as been considered
in Ref.~\onlinecite{circular_scattering}.) The potential for this
problem can be defined by 
\begin{equation}
 V_C(r) = \left\{ \begin{array}{ll}
0 & \mbox{for $r<R$} \\
\infty & \mbox{for $r\geq R$}
\end{array}
\right.
\,.
\end{equation}
The (unnormalized) solutions of the corresponding 2D Schr\"odinger
equation are given by
\begin{equation}
 u_{(m)}(r,\theta) = J_{|m|}(kr) e^{im\theta},
\end{equation}
where the quantized angular momentum values are given by $L_z = m
\hbar$ for $m=0, \pm 1, \pm 2,\ldots$ and the $J_{|m|}(kr)$ are the
(regular) Bessel functions of order $|m|$. 

The wavenumber, $k$, is related to the energy via 
$k = \sqrt{2\mu E/\hbar^2}$ and the energy eigenvalues are
quantized by the boundary conditions at the infinite wall
at $r=R$, namely $J_{|m|}(z=kR) =0$. The quantized energies are
then given by
\begin{equation}
E_{(m,n_r)} = \frac{\hbar^2 [z_{(m,n_r)}]^2}{2\mu R^2},
\end{equation}
where $z_{(m,n_r)}$ denotes the zeros of the Bessel function of
order
$|m|$, and $n_r$ counts the number of radial nodes. 

The energy spectrum is doubly degenerate for $|m| \neq 0$
corresponding to the equivalence of clockwise and counter-clockwise
($m>0$ and $m<0$) motion. We therefore see a pattern of
degeneracies very similar to that of both the square and
equilateral triangle wells, with two equal energy states for each
$(|m|>0,n_r)$ value, and a single one for each $(m=0,n_r)$. Because
the quantum number dependence of the energy eigenvalues is the
determining factor in the structure of wave packet revivals, we
need to examine the $m,n_r$ dependence of $E_{(m,n_r)} \propto
[z_{(m,n_r)}]^2$.

As a first approximation, we can look at the large $z$ behavior of
the Bessel function solutions\cite{math_handbook} for fixed values
of
$|m|$, namely
\begin{equation}
J_{|m|}(z) 
\longrightarrow 
\sqrt{\frac{2}{\pi z}}
\cos\left(z - \frac{|m|}{2}- \frac{\pi}{4}\right) + \cdots
\end{equation}
With this approximation, the zeros would be given by
$z_{(m,n_r)} -\frac{|m|}{2} - \frac{\pi}{4} \approx
(2n_r+1)\frac{\pi}{2}$,
or
\begin{equation}
z_{(m,n_r)} \approx \bigl(n_r+\frac{|m|}{2}+\frac{3}{4}\bigr)\pi .
\label{first_approximation}
\end{equation}
If this result were exact, the quantized energies would depend
quadratically on two quantum numbers, and there
would be exact wave packet revivals, just as for the square or
equilateral triangle billiards. 
However,
there are important corrections to this first-order result, which
means that the Bessel function zeros are not given by exact
integral values. So although there are not exact quantum revivals
in the general case, approximate revivals are present, most
especially for zero-momentum, central ($(x_0,y_0) = (0,0)$) wave
packets (constructed from purely $m=0$ states). We refer the
interested reader to Ref.~\onlinecite{robinett_pra} for details.

We can, however, examine the short-time dependence
leading to classical closed orbits by making use of the WKB
approximation to evaluate the quantized energies and their $m$
and $n_r$ dependence. If we first quantize the angular variable to
find that the angular momentum is given by $L_z = m\hbar$, we note
that in the radial direction the particle moves freely up to the
infinite wall at $r=R$, but is subject to an effective centrifugal
potential given by $V_{\rm eff}(r) = L_z^2/2\mu r^2 =
(m\hbar)^2/2\mu r^2$. A classical particle cannot penetrate this
centrifugal barrier and therefore has an inner radius (distance
of closest approach) given by $V_{\rm eff}(R_{\rm min}) =
m^2\hbar^2/(2\mu R_{\rm min}^2) = E$,
or
\begin{equation}
\label{73}
R_{\rm min} = \frac{|m|\hbar}{\sqrt{2\mu E}}
\,.
\end{equation}
We can also write Eq.~(\ref{73}) in the useful form
$R_{\rm min} = |m|R/z$ and
$E \equiv \hbar^2 z^2/(2\mu R^2)$
in terms of the dimensionless 
variable, $z$. 

The WKB quantization condition on the radial variable $r$
is then given by
\begin{eqnarray}
\int_{R_{\rm min}}^{R} k_r(r)\,dr = (n_r + C_L + C_R) \pi, \\
\noalign{\noindent where}
k_r(r) \equiv \sqrt{\frac{2\mu E}{\hbar^2}} \sqrt{1 - \frac{R_{\rm
min}^2}{r^2}}.
\end{eqnarray}
The matching coefficients\cite{wkb_approximation} are given by 
$C_L = 1/4$ and $C_R = 1/2$ which are appropriate for linear
boundaries (at the inner centrifugal barrier) and hard or
infinite wall boundaries (such as at $r=R$), respectively.
(See Appendix~A for more discussion of WKB methods.) The WKB
energy quantization condition for the quantized energies in
terms of $n_r$ explicitly and $|m|$ implicitly can then be written
in the form 
\begin{equation}
\sqrt{\frac{2\mu}{\hbar^2}} \!
\int_{R_{\rm min}}^{R} \sqrt{E - \frac{m^2 \hbar^2}{2\mu r^2}} \,dr
= 
\sqrt{\frac{2\mu E}{\hbar^2}} 
\int_{R_{\rm min}}^{R} \sqrt{1 - \frac{R_{\rm min}^2}{r^2}} \,dr
= (n_r + 3/4)\pi,
\label{wkb_condition}
\end{equation}
which defines the quantized energies, $E = E_{(m,n_r)}$,
implicitly in terms of $m,n_r$. To obtain the partial derivatives
that are necessary to evaluate the classical periods in
Eq.~(\ref{two_classical_periods}), we can then differentiate this 
condition with respect to both quantum numbers to obtain
\begin{subequations}
\begin{eqnarray}
\sqrt{\frac{\mu}{2\hbar^2}}
\left[ \int_{R_{\rm min}}^{R} \frac{dr}{\sqrt{E - m^2\hbar^2/2\mu
r^2}} \right]
\left(\frac{\partial E}{\partial n_r} \right) & = & \pi \\
\sqrt{\frac{\mu}{2\hbar^2}}
\left[ \int_{R_{\rm min}}^{R} \frac{dr}{\sqrt{E - m^2\hbar^2/2\mu
r^2}} 
\left(\frac{\partial E}{\partial m} 
- \frac{|m|\hbar^2}{\mu r^2}\right) \right]
& = & 0.
\end{eqnarray}
\end{subequations}
The condition for periodic orbits from
Eq.~(\ref{2d_closed_orbit_condition}) can then be written as
\begin{equation}
\frac{q}{p} = \frac{T_{\rm cl}^{(n_r)}}{T_{\rm cl}^{(m)}}
= \frac{|\partial E/ \partial m|}{|\partial E/\partial n_r|}
= \left(\frac{|m|\hbar}{\pi \sqrt{2\mu E}}
\right)\left[\int_{R_{\rm min}}^{R}
\frac{dr}{r\sqrt{r^2 - R_{\rm min}^2}}\right].
\end{equation}
If we evaluate the integral and use $R_{\rm min} \equiv
|m|\hbar/\sqrt{2\mu E}$, we find that
\begin{subequations}
\label{circular_closed_orbits}
\begin{eqnarray}
\frac{q}{p} &=&\frac{1}{\pi} \sec^{-1}\bigl(\frac{R}{R_{\rm
min}}\bigr)\\
\noalign{\noindent or}
R_{\rm min}(p,q) &\equiv& R_{\rm min} = R \cos\left(\frac{\pi
q}{p}\right)
\end{eqnarray}
\end{subequations}
as the condition on periodic orbits. 
To find the classical period for such closed orbits, we note that
\begin{equation}
\label{circular_closed_orbit_periods}
T_{\rm cl}^{\rm (po)} = p T_{\rm cl}^{(n_r)} =
p \bigl[ \frac{2\pi \hbar }{|\partial E/\partial n_r|} \bigr]
 = \bigl(2p \sqrt{R^2 - R_{\rm min}^2} \bigr)
\sqrt{\frac{\mu}{2E}}
= \frac{ 2pR\sin(\pi q/p) }{v_0} = \frac{L(p,q)}{v_0},
\end{equation}
where we identify $v_0 = \sqrt{2E/\mu}$ with the classical speed.
The condition in Eq.~(\ref{circular_closed_orbits}) and the
resulting path lengths/periods in 
Eq.~(\ref{circular_closed_orbit_periods})
can be obtained from strictly
geometrical arguments.\cite{robinett_po_theory,balian_and_bloch}
Just as for the square and equilateral triangle cases,
information on the classical closed orbits is clearly encoded in
the energy eigenvalue spectrum for the circular well using
Eqs.~(\ref{two_classical_periods}) and
(\ref{2d_closed_orbit_condition}). The values of these path
lengths, $L(p,q) = 2pR\sin(\pi q/p)$, and the corresponding
distances of closest approach, $R_{\rm min}/R = R\cos(\pi q/p)$,
for many of the lowest-lying closed or periodic orbits in the
circular well are collected in Table~II.

\subsection{Wave packet construction in the circular billiard}
To examine the construction of Gaussian wave packets in this
geometry, we note that any arbitrary initial 
wave packet, $\psi(r,\theta)$, in the circular billiard can be
expanded in the normalized eigenstates of the form
\begin{equation}
w_{(m,n_r)}(r,\theta) = [N_{(m,n_r)}J_{|m|}(k_{(m,n_r)}r)]
 \bigl(\frac{1}{\sqrt{2\pi}}
e^{im\theta}\bigr),
\end{equation}
where
\begin{equation}
\left[N_{(m,n_r)}\right]^2 \!\int_{0}^{R}\!
r\,[J_{|m|}(kr)]^2\,dr = 1,
\end{equation}
with expansion coefficients given by
\begin{equation}
a_{(m,n_r)} = \langle \psi(r,\theta;t=0) | w_{(m,n_r)} \rangle,
\end{equation}
which satisfy
\begin{equation}
\sum_{m=-\infty}^{+\infty}\sum_{n_r=0}^{\infty}
|a_{(m,n_r)}|^2 = 1\,. 
\label{normalization}
\end{equation}
Because the eigenstates are no longer simple trigonometric functions,
the required normalization and overlap integrals must be done
numerically, making the process of wavepacket construction much
more involved. The expectation value of the energy in this
potential well is given by 
\begin{equation}
\langle \hat{E} \rangle 
= 
\sum_{m=-\infty}^{+\infty}\sum_{n_r=0}^{\infty}
|a_{(m,n_r)}|^2 E_{(m,n_r)}
=
\sum_{m=-\infty}^{+\infty}\sum_{n_r=0}^{\infty}
|a_{(m,n_r)}|^2 \left(\frac{\hbar^2[z_{(m,n_r)}]^2}{2\mu
R^2}\right),
\label{general_energy}
\end{equation}
and this constraint is another useful one for numerical checks. 

In this more symmetrical geometry, we also can evaluate
expectation values of powers of the angular momentum,
$\hat{L}_z^k$, which is the other important conserved quantity. 
We find that
\begin{equation}
\langle \hat{L}_z^{k} \rangle
= 
\sum_{m=-\infty}^{+\infty} 
\sum_{n_r = 0}^{\infty} |a_{(m,n_r)}|^2 (m\hbar)^{k}
\,. 
\label{general_angular_momentum}
\end{equation}
The subsequent time-dependence of any such wave packet is then given by
\begin{equation}
\psi(r,\theta;t) = \sum_{m=-\infty}^{+\infty} \sum_{n_r=0}^{\infty}
a_{(m,n_r)} w_{(m,n_r)}(r,\theta) \, e^{-iE_{(m,n_r)}t/\hbar},
\end{equation}
and the autocorrelation function
is given by\cite{nauenberg}
\begin{equation}
A(t) \equiv 
\langle \psi(r,\theta;t) | \psi(r,\theta,0) \rangle
=
\sum_{m=-\infty}^{+\infty} \sum_{n_r=0}^{\infty} 
|a_{(m,n_r)}|^2 e^{-iE_{(m,n_r)}t/\hbar}
\,. 
\label{circular_autocorrelation_function}
\end{equation}

We now focus on the specific Gaussian form in 
Eq.~(\ref{initial_gaussian}), besides the expectation values
calculated above for such a form. We can explicitly evaluate
the conserved powers of the angular momentum, namely
\begin{equation}
\langle \hat{L}_z \rangle=
\langle x \hat{p}_y - y \hat{p}_x \rangle
= \langle x \rangle \langle \hat{p}_y \rangle
- \langle y \rangle \langle \hat{p}_x \rangle
= x_0p_{0y} - y_0 p_{0x},
\label{gaussian_angular_momentum_1}
\end{equation}
and
\begin{equation}
\langle \hat{L}_z^2 \rangle
= (x_0p_{0y} - y_0 p_{0x})^2
+ \frac{b^2}{2} \left[(p_{0x})^2 + (p_{0y})^2\right]
+ \frac{\hbar^2}{2b^2} \left[(x_0)^2 + (y_0)^2\right],
\label{gaussian_angular_momentum_2}
\end{equation}
so that the spread (or uncertainty) in the angular momentum is
given by
\begin{equation}
(\Delta m )\hbar 
\equiv 
\Delta L_z = 
\sqrt{
\frac{b^2}{2} \left[(p_{0x})^2 + (p_{0y})^2\right]
+ \frac{\hbar^2}{2b^2} \left[(x_0)^2 + (y_0)^2\right]
}.
\label{gaussian_angular_momentum_spread}
\end{equation}
This form can be understood intuitively from a
simple error propagation argument that would give the error
in the product
$L \sim r p$ to be 
$(\Delta L/L)^2 
= (\Delta r/r)^2
+
(\Delta p/p)^2$
or
\begin{equation}
\Delta L = \sqrt{ p^2 (\Delta r)^2 + r^2 (\Delta p)^2}
\,.
\label{classical_argument}
\end{equation}
We note that even for wavepackets for which the expectation value
of the angular momentum vanishes, it may still be necessary to
include
$m \neq 0$ components; for example, a wave packet with $(x_0,y_0)
= (0,0)$ and $(p_{0x},p_{0y}) = (0,p_0)$, will have a necessary
spread in angular momentum values given by
$\Delta m = bp_0/\sqrt{2}\hbar$, which increases linearly with 
the initial momentum. For such cases, because $\langle \hat{L}_z
\rangle = 0$, we must have relations such as $|a_{(m,n_r)}|
= |a_{(-m,n_r)}|$ to ensure that
Eq.~(\ref{general_angular_momentum}) is satisfied.

Because of the required 
numerical evaluation of the expansion coefficients, 
it is useful to be able to compare the general results for 
$\langle E \rangle$ and $\langle \hat{L}_z^{(1,2)} \rangle$ in 
Eqs.~(\ref{general_energy}) and (\ref{general_angular_momentum}) 
with the specific results for the Gaussian in
Eqs.~(\ref{gaussian_energy}), (\ref{gaussian_angular_momentum_1}),
and (\ref{gaussian_angular_momentum_2}).
As an example of such an expansion, we have numerically evaluated
the $a_{(m,n_r)}$ for Gaussian wave packets characterized by
$y_0, p_{0x} = 0$, $p_{0y} = 100$, and $x_0/R= 0.0,\, -0.5$ and
0.70, that is, with the same total (kinetic) energy, but
different values of the angular momentum;
 the same physical parameters of Eq.~(\ref{physical_parameters})
are used, along with $R=1$. The various expansion probabilities,
$|a_{(m,n_r)}|^2$, are sorted by magnitude, and then summed to
confirm that the numerical evaluation has captured all of the
probability density. We can then illustrate which regions of the
$(m,n_r)$ space are most populated in the construction of each wave
packet and the results are shown in Fig.~5.
In each case, the (approximately elliptical) inner shaded areas
correspond to regions in $(m, n_r)$ space containing 68\%
(inner dark region) of the total probability, while the outer dark
regions surround 99.7\% of the $\sum_{(m,n_r)} |a_{(m,n_r)}|^2$.
We note that the required spread in $m$ values increases
for those states with 
increasing values of $|\langle \hat{L} \rangle| = x_0 p_{0y}$.

We can also follow the short-time development of Gaussian wave
packets in the circular billiard for various initial conditions 
to look for evidence of the classical periodicities described by 
Eqs.~(\ref{circular_closed_orbits}) and
(\ref{circular_closed_orbit_periods}). In this case we construct
packets with $y_0=0$, $p_{0x} =0$, $p_{0y} = 100$ and various
values of $x_0 = R_{\rm min}$. (We use a much smaller value of
$p_{0}$ than in earlier calculations for the square well because we
are restricted to evaluating the expansion coefficients
numerically which becomes prohibitively time-consuming for larger
values: the repeated evaluation of high-order Bessel functions is
numerically intensive in the programs we use and is the limiting
factor. This unavoidable restriction
means that the spreading time,
$t_0$, is of the same order as the classical periods shown.) We
plot in Fig.~6 the autocorrelation function
$|A(t)|^2$ versus $t/\tau$, where $\tau \equiv R/v_0$. We
note that isolated features in $A(t)$ are present at the values of
$T_{\rm cl}^{\rm (po)}/\tau = 2p\sin(\pi q/p)$ and $R_{\rm min}/R =
\cos(\pi q/p)$ values given in Table~II, corresponding to classical
closed orbits in this geometry. 

We can extend this analysis to the half-circle billiard,
obtained by folding across a diameter in the same way as we have
done for the half-square and half-equilateral triangle. In this
case, we use a single copy of the $m > 0$
eigenstates.\cite{robinett_pra}

\section{Concluding remarks}\label{remarks}
We have examined the quantum mechanical time-development of
localized (Gaussian-like) wave packets in three billiard systems,
with square, equilateral triangle, and circular footprints. We
have used the well-known (and not-so-well-known) energy eigenvalue
spectrum in each case to discuss the classical periodicity and
quantum revival time scales in such systems using the
autocorrelation function as a probe. In the first two cases, we
used the corresponding energy eigenfunctions to calculate
closed form, analytical approximations for the expansion
coefficients of the Gaussian wave packets, suitable for fast
numerical evaluations for arbitrarily large momentum values,
provided the initial wave packets are well-localized within the
well and reasonably far from any infinite wall boundary. For the
circular well, we have performed a similar analysis, using a WKB
analysis to demonstrate the expected short-time classical
periodicities and a numerical approach to evaluate $A(t)$.

In each case, we have also been able to consider half-well
geometries, obtained by folding of the original footprint
along an obvious axis of symmetry for which solutions were
available as a subset of those for the full-well. 

Further extensions of these studies could include more detailed 
examinations of rectangular geometries, or even
three-dimensional (3D) parallelepipeds. The case of an annular
billiard (with both outer and inner infinite
walls)\cite{other_annular_ones} can be handled in much the same
way as described here for the 2D circular well, and the same WKB
approach can be used to extract the pattern of allowed classical
closed orbits, which is much richer than in the simple circular
billiard.

The spherical billiard also can be discussed with
many of the same techniques. The energy eigenstates now consist of
products of spherical harmonics, $Y_{l,m}(\theta,\phi)$, instead of
the simple
$e^{im\theta}$ angular states, with spherical Bessel functions for the
radial component. We might expect that the resulting change in the
angular momentum term from two to three dimensions would be given
by the substitution
$m^2 \rightarrow l(l+1)$, but it turns out that for a WKB analysis
it is more appropriate\cite{wkb_approximation} to use $(l+1/2)^2$
instead of 
$l(l+1)$ (the so-called Langer modification.\cite{langer_ref})
This identification
 still gives the same result for the
classical periods, which is not surprising because angular
momentum conservation implies that all the allowed 3D classical
orbits will be planar and reduce to the 2D case. This effect can
be seen rather generally in 3D central potential systems where we
know that the quantized energy eigenvalues do not depend on the
$m$ quantum number, namely the $E(n_r,l,m) = E(n_r,l)$ are 
functions of $n_r$ and $l$ alone. This independence implies that
the classical periodicity corresponding to the $m$ quantum number
in Eq.~(\ref{two_classical_periods}) will be $T_{\rm cl}^{(m)} =
2\pi
\hbar/|\partial E/\partial m| \rightarrow
\infty$ which becomes irrelevant, leaving only $T_{\rm cl}^{(n_r)}$
and $T_{\rm cl}^{(l)}$ to beat against each other to provide
closed orbits.

In the context of the spherical billiard, we note that the
spherical Bessel functions, $j_{l}(z)$, can be related to the ones
considered here for the 2D case via
$j_{l}(z) = \sqrt{\pi/2 z} J_{l+1/2}(z)$. For purely
central (no angular momentum, $s$-wave) Gaussian wave packets
(therefore with initial position at $(x_0,y_0,z_0) = (0,0,0)$ and
with vanishing momentum), the required $l = 0$ spherical Bessel
function is proportional to
$\sin(z)/z$, which {\it does} have exact integral values of $z=kR
= n\pi$. For this case only, there will be exact quantum revivals
in the spherical billiard.

The solutions for the equilateral triangle case
($N=3$ regular polygon) form
a subset of those required for a complete study of the ($N=6$)
hexagonal billiard and that case is currently under investigation.
It is an interesting and open question as to
whether there are other $N$ values (besides $N=3$ and 4)
for which there exist exact quantum revivals as well as well as
how the
$N\rightarrow
\infty$ limit of the circular billiard (where only approximate
revivals are present) is reached. Approximation methods
appropriate for such quantum billiard systems
\cite{approximation} may be useful in answering such questions.

We also can extend the notions explored here dealing with
classical periodicity to non-billiard systems. For example,
the extension of Eqs.~(\ref{two_classical_periods}) and
(\ref{2d_closed_orbit_condition}) to three dimensions is obvious.
As mentioned, for central potentials the $T_{\rm cl}^{(m)}$
classical period is irrelevant, so that closed or periodic orbits
are the result of the condition
$p T_{\rm cl}^{(n_r)} = T_{\rm cl}^{\rm (po)} = qT_{\rm cl}^{(l)}$ 
for the $(n_r,l)$ quantum numbers.
It is a familiar result of classical mechanics (Bertrand's
theorem\cite{rosner,bertrand}) that for the 3D power law
potentials ($V(r) \propto r^{k}$), only the harmonic oscillator
($k=2$) and Coulomb potentials ($k=-1$) give rise to closed orbits
for all bound state trajectories.\cite{rosner} It is 
straightforward to exhibit a similar statement for the quantum
versions of these systems in terms of their expected classical
periods, $T_{\rm cl}^{(po)}$.

\begin{acknowledgments}
The work of R.\ W.\ R.\ was supported, in part, by the National
Science Foundation under Grant DUE-9950702. The work of M.\ A.\ D.\
was supported, in part, by the Commonwealth College of the
Pennsylvania State University under a Research Development Grant
(RDG).
\end{acknowledgments}

\appendix

\section{WKB analysis of classical periods}
As a comparison with the formalism leading to
Eq.~(\ref{classical_period_definition}), an alternative derivation of the
expression for the classical periodicity of a bound state system in
terms of its quantized energy levels can be obtained from a simple
WKB argument. For a particle of fixed energy E in a bound state
potential, $V(x)$, we have
$E = \mu v(x)^2/2 + V(x)$ 
and the short time, $dt$, required to traverse a distance $dx$ 
can be obtained from this energy conservation connection
and integrated over a single cycle to obtain the classical period via
\begin{subequations}
\begin{eqnarray}
dt = \sqrt{\frac{\mu}{2}} \frac{dx}{\sqrt{E-V(x)}},\\
\noalign{\noindent or}
\tau = 2\int_{a}^{b}\,dt =
2 \sqrt{\frac{\mu}{2}}\int_{a}^{b} \frac{dx}{\sqrt{E-V(x)}}.
\label{classical_period_wkb}
\end{eqnarray}
\end{subequations}
The WKB quantization condition for this potential 
can be written in the form\cite{wkb_approximation} 
\begin{equation}
\label{a2}
\sqrt{2\mu} \int_{a}^{b} \sqrt{E_n -V(x)}\, dx
 = (n +C_{L} + C_{R}) \pi \hbar 
\end{equation}
in terms of the matching coefficients $C_L$ and $C_R$.
Recall that
$C_{L,R} = 1/4$ at linear or smooth turning points where the
functions are matched smoothly onto Airy solutions, 
while $C_{L,R} = 1/2$ at infinite wall type boundaries
where the wavefunction must vanish.\cite{wkb_approximation} This
difference is sometimes described as being due to the fact that
WKB wavefunctions can penetrate roughly one-eighth of a
wavelength into the classically disallowed region, provided the
barrier is not infinitely high.\cite{saxon}

Equation~(\ref{a2}) can be differentiated implicitly
with respect to the quantum number $n$ to obtain
\begin{equation}
\label{a3}
\sqrt{2\mu}\! \int_{a}^{b} \!
\frac{|dE_n/dn|\,dx}{2\sqrt{E_n-V(x)}} = 
\pi \hbar .
\end{equation}
Equation~(\ref{a3}) can be related to the classical period
in Eq.~(\ref{classical_period_wkb}) to give
\begin{equation}
\tau_n \equiv \sqrt{2\mu }\int_{a}^{b} (E_n - V(x))^{-1/2}\,dx
= \frac{2\pi \hbar}{|dE_n/dn|},
\label{classical_period_from_wkb}
\end{equation}
as in Eq.~(\ref{classical_period_definition}).
The most obvious example of such a connection is for the harmonic
oscillator, where the WKB condition gives the exact eigenvalues,
$E_n = (n+1/2) \hbar
\omega$, and the classical period from
Eq.~(\ref{classical_period_from_wkb}) is
\begin{equation}
\tau_{n} = 
\frac{2\pi \hbar}{|dE/dn|} =
\frac{2\pi \hbar}{\hbar \omega} = \frac{2\pi}{\omega}
\end{equation}
as expected. 

\section{Expansion coefficients for the 1D infinite well 
in momentum space}

Although it is typical to discuss the expansion coefficients for a
Gaussian wave packet in the 1D infinite well in position space, as
in Eqs.~(\ref{expansion_in_eigenstates}) or 
(\ref{approximate_expansion_coefficients}), it is instructive
to examine the same problem in momentum space. As mentioned in
Sec.~\ref{sec3}, the requirement of the necessary matching of
the oscillatory $e^{-ip_x x/\hbar}$ factor
in $\psi_{G}(x,0)$ with the corresponding $e^{\pm in \pi x/a}$
factors from the bound state $u_n(x)$ is contained explicitly in
the
$\exp(-b^2(p_0 \pm n \pi \hbar/a)^2/2\hbar^2)$ terms in 
Eq.~(\ref{approximate_expansion_coefficients}). 
There must also be an appropriate overlap
between the envelope of
$\psi_{G}(x,0)$ contained in the $\exp(-(x-x_0)^2/2b^2)$ factor
and the extent of the bound state wavefunctions. That is, it must
be inside the well, as illustrated in Fig.~1. 

The form of the momentum-matching can be qualitatively understood
in the following way. The oscillatory parts of the integrals
required for the evaluation of $a_n$ will contain factors such as
$\exp(i(p_0 \pm n\pi \hbar/a)x/\hbar)$ to be integrated over $x$.
An integral over all space would give rise to a
$\delta$-function term of the form $\delta([p_0\pm n \pi
\hbar/a]/\hbar)$, because of the cancellations arising from the
rapid oscillations of the integrand. However, the integrals are
effectively cut off by the spatial extent of the initial wave
packet, a width of order $\Delta x \sim b$, which instead gives a
sharply peaked function of 
$|p_0 \pm n\pi \hbar/a|b/\hbar$, as seen in
Eq.~(\ref{approximate_expansion_coefficients}).

The same information must be encoded in a
momentum space representation of the evaluation of the $a_n$
expansion coefficients. The required Fourier transform 
\begin{equation}
\phi_{n}(p) = \frac{1}{\sqrt{2\pi \hbar}}
\int_{0}^{a}\, u_{n}(x) \, e^{-ipx/\hbar}\,dx
\end{equation}
of the bound-state wavefunction
$u_n(x) = \sqrt{\frac{2}{a}}\sin(\frac{n \pi x}{a})$
is given by
\begin{eqnarray}
\phi_{n}(p) & = & - \frac{a}{\sqrt{4\pi \hbar a}}
\left[ 
\frac{e^{i(n\pi - pa/\hbar)} -1}{(n\pi - pa/\hbar)}
+
\frac{e^{-i(n\pi + pa/\hbar)} -1}{(n\pi + pa/\hbar)}
\right] \\
& = & - \frac{ai}{\sqrt{4\pi \hbar a}}
\left[
e^{in\pi} \frac{\sin[(n\pi - pa/\hbar)/2]}{[(n\pi - pa/\hbar)/2]}
-
e^{-in\pi} \frac{\sin[(n\pi + pa/\hbar)/2]}{[(n\pi + pa/\hbar)/2]}
\right] e^{-ipa/2\hbar}, \nonumber 
\end{eqnarray}
while the initial Gaussian from 
Eq.~(\ref{gaussian_wave_packet_momentum_space}) 
is
\begin{equation}
\phi_{G}(p) = \sqrt{\frac{\alpha}{\sqrt{\pi}}}
\, e^{-\alpha^2(p-p_0)^2/2} \, e^{-ipx_0/\hbar}
\,.
\end{equation}
The (exact) expansion coefficients now have the form
\begin{equation}
a_{n} = \int_{-\infty}^{+\infty} \phi_{n}^{*}(p)\,
\phi_{G}(p,0)\,dp
\label{momentum_space_expansion_coefficient}
\end{equation}
The matching in momentum space now comes from the requirement
that the peak in $\phi_{G}(p)$ (near $p_0$) must match with
those in $\phi_{n}(p)$ (near $p = \pm n \pi \hbar/a$). The
information on whether the initial wave packet is contained inside
the $(0,a)$ infinite well now arises from the phase information
contained in the oscillatory parts of the momentum space versions:
specifically, the $a_{n}$ integral in
Eq.~(\ref{momentum_space_expansion_coefficient}) contains terms
such as
\begin{equation}
[\phi_{n}^{*}(p)]
[\phi_{G}(p,0)]
\propto
[e^{-ipa/2\hbar}]^{*}[e^{-ipx_0/\hbar}]
= e^{ip/\hbar(a/2-x_0)},
\label{momentum_space_phase_factor}
\end{equation}
which, if integrated by itself over all $p$-space, would give rise
to a 
$\delta(x_0-a/2)$ term. However, this singular behavior
is again
softened to be a sharply peaked function of $|x_0-a/2|$, due
to the finite extent of the $p$-integrals. It should not be
surprising that the roles played by the envelope/overlap
requirement versus that played by the oscillatory information
are complementary in the
$x$- versus
$p$-space descriptions of the integrals in
Eqs.~(\ref{position_space_expansion_coefficient}) and 
(\ref{momentum_space_expansion_coefficient})

We can easily generalize this last result to the case of an
infinite well of width $a$, but located at an arbitrary location
along the 1D axis defined over the range $(d,d+a)$. In
that case, the energies are unchanged, and the position space
eigenfunctions are simply shifted to
\begin{equation}
\tilde{u}_{n}(x) = 
u_n(x-d) = \sqrt{\frac{2}{a}}\sin\bigl(\frac{n \pi
(x-d)}{a}\bigr).
\end{equation}
The Fourier transform to obtain the corresponding momentum space
eigenstates goes through as before with a simple change of
integration variables, $x-d \rightarrow y$, giving
\begin{equation}
\tilde{\phi}_{n}(p) = \phi_{n}(p) e^{-ipd/\hbar},
\end{equation}
in which case the appropriate phase factor, analogous to that
in Eq.~(\ref{momentum_space_phase_factor}), is
$\exp(ip/\hbar(d+a/2-x_0))$; values of $x_0$ beyond the
boundaries of the new well lead to exponential suppression.

\section{Problems}
We offer some suggested problems that are motivated by the
suggestions in Sec.~\ref{remarks}.

\smallskip \noindent {\bf Problem 1}. Extend the results of
Sec.~\ref{sec4b} for a square to a rectangular billiard with
sides $L_x
\times L_y$. Calculate the allowed energies and the classical
periods. Discuss the patterns of exact and/or fractional
revivals. (See Refs.~\onlinecite{bluhm_2d} and
\onlinecite{other_2d} for details.)

\smallskip\noindent{\bf Problem 2}. Extend the relationship in
Eq.~(\ref{2d_closed_orbit_condition}) to a system with three
quantum numbers to find the allowed classical periods. Apply this
extension of Eq.~(\ref{2d_closed_orbit_condition}) 
to the system of a cubical box of side $L$,
generalizing the result of Eq.~(\ref{square_box_relation}).

\smallskip\noindent {\bf Problem 3}. Visualize some of the closed
orbits for the equilateral triangle geometry. (See
Ref.~\onlinecite{robinett_annals} for examples.)

\noindent{\bf Problem 4}. Generalize the WKB results of
Sec.~\ref{sec:closed} to show how the closed orbits (periods and
path lengths) in an annular infinite well are obtained. How is
Eq.~(\ref{wkb_condition}) changed, including the limits of
integration and the matching coefficients,
$C_L$ and $C_R$? (See Ref.~\onlinecite{other_annular_ones} for
details.)

\smallskip\noindent
{\bf Problem 5}. Plot the equivalent of Fig.~5 for Gaussian wave
packets in a square well, initially centered at the origin, with
initial momentum components $(p_{0x},p_{0y}) = (p_0
\cos(\theta),p_0\sin(\theta))$ for various values of $\theta$.
That is, evaluate the expansion coefficients,
$a_{(n_x,n_y)} = a_{n_x} a_{n_y}$, and the corresponding
probabilities, and plot where the probability is distributed in
the $(n_x,n_y)$ plane.

\newpage

\section*{Tables}

\begin{center}
\begin{tabular}{|r|c|c||r|c|c|} \hline
$\theta$ (deg) & period $T_{\rm cl}^{\rm (po)}/\tau$ & $(p,q)$ 
&
$\theta$ (deg) & period $T_{\rm cl}^{\rm (po)}/\tau$ & $(p,q)$ \\ 
\hline
$0.00$ & $1.00, 2.00, 3.00,\ldots $ & $(1,0)$ & $23.96$ & $9.85$ &
$(9,4)$ \\ \hline
$6.34$ & $9.06$ & $(9,1)$ & $26.57$ & $2.24, 4.47, 6.71, 8.94$ & $(2,1)$ \\ \hline
$7.13$ & $8.06$ & $(8,1)$ & $29.74$ & $8.06$ & $(5,3)$ \\ \hline
$8.13$ & $7.07$ & $(7,1)$ & $30.96$ & $5.83$ & $(7,4)$ \\ \hline
$9.46$ & $6.08$ & $(6,1)$ & $32.00$ & $9.43$ & $(8,5)$ \\ \hline
$11.31$ & $5.10$ & $(5,1)$ & $33.69$ & $3.61, 7.21$ & $(3,2)$ \\ \hline
$12.53$ & $9.22$ & $(9,2$) & $35.54$ & $8.60$ & $(7,5)$ \\ \hline
$14.04$ & $4.12, 8.25$ & $(4,1)$ & $36.87$ & $5.00, 10.00$ & $(4,3)$ \\ \hline
$15.95$ & $7.28$ & $(7,2)$ & $38.66$ & $6.40$ & $(5,4)$ \\ \hline
$18.43$ & $3.16, 6.32, 9.49$ & $(3,1)$ & $39.81$ & $7.81$ & $(6,5)$ \\ \hline
$20.56$ & $8.54$ & $(8,3)$ & $40.60$ & $9.22$ & $(7,6)$ \\ \hline
$21.80$ & $5.38$ & $(5,2)$ & $45.00$ & $1.41, 2.83, 4.24, \ldots$
& $(1,1)$ \\ \hline
$23.20$ & $7.62$ & $(7,3)$ & & & \\ \hline
\end{tabular}
\end{center}

\vskip 1cm
\noindent
{\bf Table~I}: The periods for the classical closed orbits, $T_{\rm
cl}^{\rm (po)}/\tau =
\sqrt{p^2+q^2}$, where $\tau \equiv 2a/v_0$, 
and the corresponding initial angles, $\tan(\theta)
= q/p$, for the square billiard.
All periodic orbits with $T_{\rm cl}^{(po)}/\tau \leq 10$ are
included. Values for $45^{\circ} < \theta< 90^{\circ}$ are the same
as those for $90^{\circ} - \theta$.

\newpage

\begin{center}
\begin{tabular}{|c|c|c||c|c|c|} \hline
$(p,q)$ & $L/R = 2p\sin(\pi q/p)$ & $R_{\rm min}/R =$ &
$(p,q)$ & $L/R = 2p\sin(\pi q/p)$ & $R_{\rm min}/R =$ \\ 
 & & $\cos(\pi q/p)$ &
 & & $\cos(\pi q/p)$ \\ \hline
$(2,1)$ & $4.00, 8.00, 12.00, 16.00, \ldots$ & $0.00$ & $(6,3)$ &
$12.00, \ldots $ & $0.00$ \\ \hline
$(3,1)$ & $5.20, 10.39, \ldots.$ & $0.50$ & $(7,3)$ & $13.65,
\ldots$ &
$0.22$ \\ \hline 
$(4,1)$ & $5.66, 11.31, \ldots.$ & $0.71$ & $(8,3)$ & $14.78,\ldots$
& $0.38$ \\ \hline
$\vdots$ & $\vdots$ & $\vdots$ & $(9,3)$ & $15.59, \ldots$ & $0.50$
\\ \hline
$(\infty,1)$ & $6.28, 12.57, \ldots$ & $1.00$ & $(10,3)$ & $16.18,
\ldots$ & $0.59$ \\ \hline & & & $(11,3)$ & $16.63, \ldots$ &
$0.66$ \\
\hline
$(4,2)$ & $8.00, 16.00,\ldots$ & $0.00$ & $(12,3)$ & $16.97,
\ldots$ &
$0.71$ \\ \hline
$(5,2)$ & $9.51, \ldots$ & $0.31$ & $(13,3)$ & $17.24,\ldots$ &
$0.75$
\\ \hline
$(6,2)$ & $10.39, \ldots$ & $0.50$ & $\vdots$ & $\vdots$ & $\vdots$
\\ \hline
$(7,2)$ & $10.95, \ldots$ & $0.62$ & $(\infty,3)$ &$ 18.85, \ldots$
&
$1.00$ \\ \hline
$\vdots$ & $\vdots$ & $\vdots$ & & & \\ \hline
$(\infty,2)$ &$ 12.57, \ldots$ & $1.00$ & & & \\ \hline
\end{tabular}
\end{center}
\vskip 1cm
\noindent
{\bf Table~II}: Tabulated values of the path length, $L/R =
2p\sin(\pi q/p)$, from Eq.~(\ref{circular_closed_orbit_periods})
and the distance of closest approach, $R_{\rm min}/R = \cos(\pi
q/p)$, from Eq.~(\ref{circular_closed_orbits}) 
for low-lying closed or periodic orbits for the circular well
characterized by $(p,q)$. All closed orbits with path lengths
satisfying $L(p,q)< 20R$ are included, as are their low-lying
recurrences.

\newpage

\begin{flushleft}
{\large {\bf Figure Captions}}
\end{flushleft}
\vskip 0.5cm

\begin{itemize}

\item[Fig.\thinspace 1.] Plots of the expansion coefficients,
$\sum_{n=1}^{\infty} |a_n|^2$ (bottom), and the average energy,
$\sum_{n=1}^{\infty} E_n |a_n|^2$ (middle), 
for zero-momentum Gaussian wave packets in the 1D infinite well, 
as the central position, $x_0$, is varied from the center ($x_0/a = 1/2$)
to beyond the edge of the well ($x_0/a > 1$). The 
expression in Eq.~(\ref{position_space_expansion_coefficient}) is
used and contributions from
the $40$ lowest-lying states are included. The solid (dashed)
curves correspond to $\Delta x_0 = 0.05$ ($0.1$). As expected, 
the normalization 
decreases uniformly from unity as more of the initial wave packet
is outside the well. The average energy, however, increases as the
initial wave packet is placed near the edge of the well, as the
expansion in eigenstates attempts to reproduce the sharp discontinuity
at the wall (as shown for the $x_0 = 0.95$ packet at the top.)

\item[Fig.\thinspace 2.] Plots of the autocorrelation function,
$|A(t)|^2$ versus $t$, over one revival time for zero-momentum 
($p_0 = 0$) Gaussian wave packets in the 1D infinite well. Plots for
different initial position values, $x_0$, from the center ($x_0/a = 0.5$)
to near one edge ($x_0/a = 0.8$) are shown. Exact revivals at 
$T_{\rm rev} = 4\mu a^2/\hbar \pi$ are obvious for all initial
positions, but special exact revivals at shorter times due to
obvious symmetries in the eigenstate expansion at $x_0 = a/2$ (for
multiples of $T_{\rm rev}/8$, one of which is highlighted in the
rectangular area) and at $x_0 = 2a/3$ (for multiples of
$T_{\rm rev}/3$, highlighted in the two elliptical areas) are also
apparent. We note that there are partial revivals for 
$x_0/a = 0.8$ (top line) for times given by $0.4T_{\rm
rev}$ and $0.6T_{\rm rev}$ (shown by the diamonds.)

\item[Fig.\thinspace 3.] Plots of the autocorrelation function,
$|A(t)|^2$ versus $t$ (left column), over one revival time for $x_0
= a/2$ wave packets with increasing values of average momentum, $p_0$.
For the bottom case (d), the classical periodicity is so short
that the individual periods cannot be resolved, in contrast to
case (c) where the periods are clearly visible. Locations of
fractional (partial) revivals are indicated by arrows for case (d)
where they are most obvious. The corresponding values of the
expansion coefficients, $|a_n|^2$ versus
$n$, are shown on the right. 

\item[Fig.\thinspace 4.] Plots of the autocorrelation function,
$|A(t)|^2$ versus $t$, for the square billiard. The plots are
over a time period to ten classical
back-and-forth periods, $10\tau$, where $\tau = 2L/v_0$.
Plots for different values of the initial angle, $\tan(\theta)
= p_{0y}/p_{0x}$ are shown. The stars indicate the positions of
classical closed orbits (and recurrences) as shown in Table~I. 

\item[Fig.\thinspace 5.] Values of the (numerically evaluated) 
zeros of the Bessel
function, $J_{|m|}(z=ka)$ as the quantum number, $m$, 
which determine the energy eigenvalues for the circular well,
is varied. The (roughly elliptical) shaded regions correspond to
regions in 
$(m,z=k_{m,n_r}$) space inside which 68\% (inner) and 99.7\%
of the $\sum_{(m,n_r)} |a_{(m,n_r)}|^2$ is contained. The three
cases shown correspond to $y_0,p_{0x} = 0$, $p_{0y} = 100$ and $x_0
= -0.5, 0.0$, and $0.7$ for Gaussian wave packets with the
physical parameters in Eq.~(\ref{physical_parameters}). The common
value of $p_{0y} = \hbar k = 100$ is shown as a horizontal dashed
line. (The faint vertical banding is an artifact of our plotting
program.)

\item[Fig.\thinspace 6.] Plots of the autocorrelation function,
$|A(t)|^2$ versus $t$, for the circular billiard. The plots are for
$x_0, p_{0y} \neq 0$ with $y_0=p_{0x} = 0$ as one increases
$x_{0}/R = R_{\rm min}/R$. The stars indicate the locations of
classical closed orbits with periods and $R_{\rm min}/R$ values
given by Eqs.~(\ref{circular_closed_orbit_periods}) and 
(\ref{circular_closed_orbits}), respectively as shown in Table~II.
Examples of closed orbit trajectories for several simple $(p,q)$ 
combinations are also shown at the corresponding values of
$R_{\rm min}/R$.

\end{itemize}


\newpage

\begin{figure}
\epsfig{file=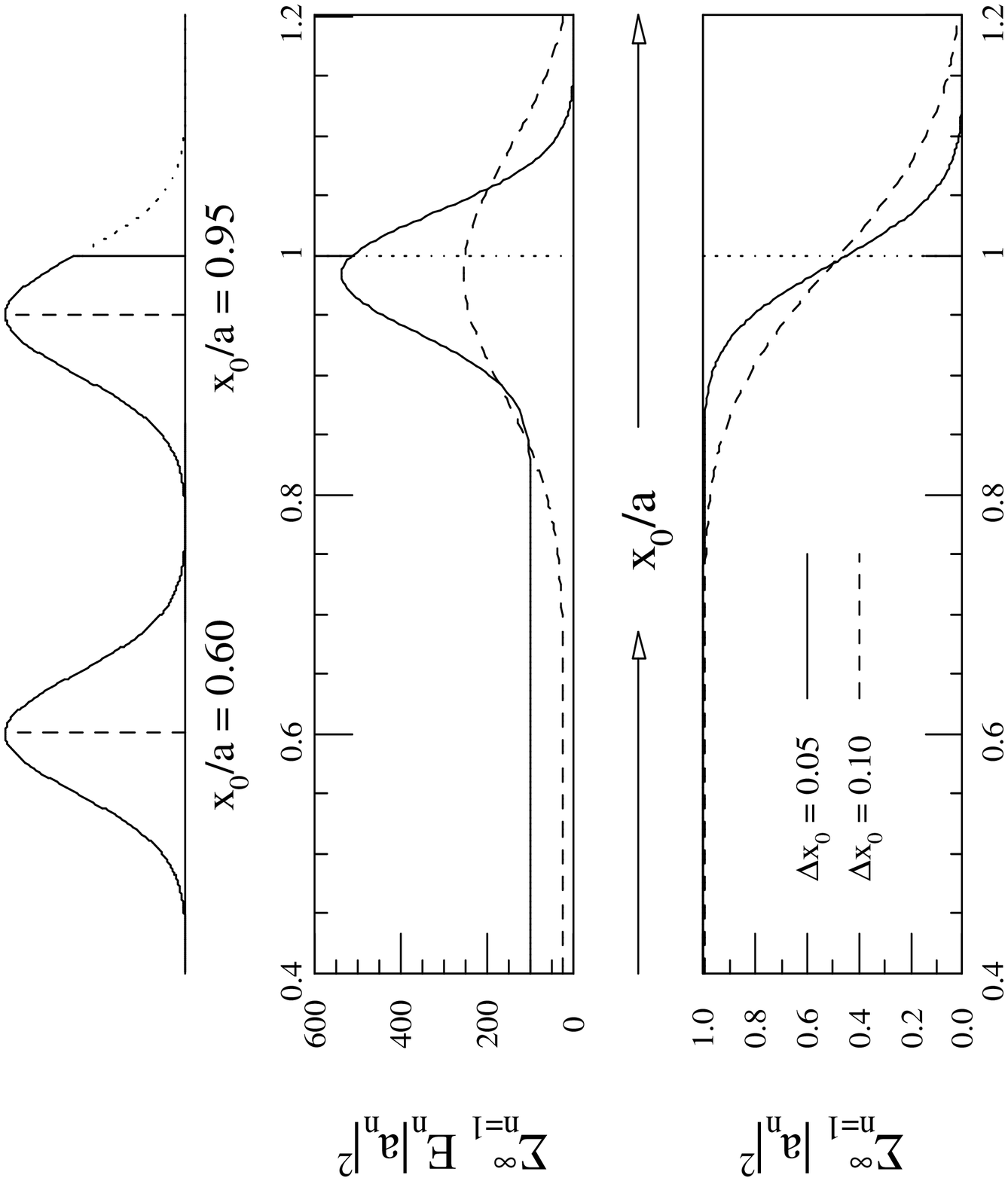,width=12cm,angle=0}
\caption{ }
\end{figure}

\newpage

\begin{figure}
\epsfig{file=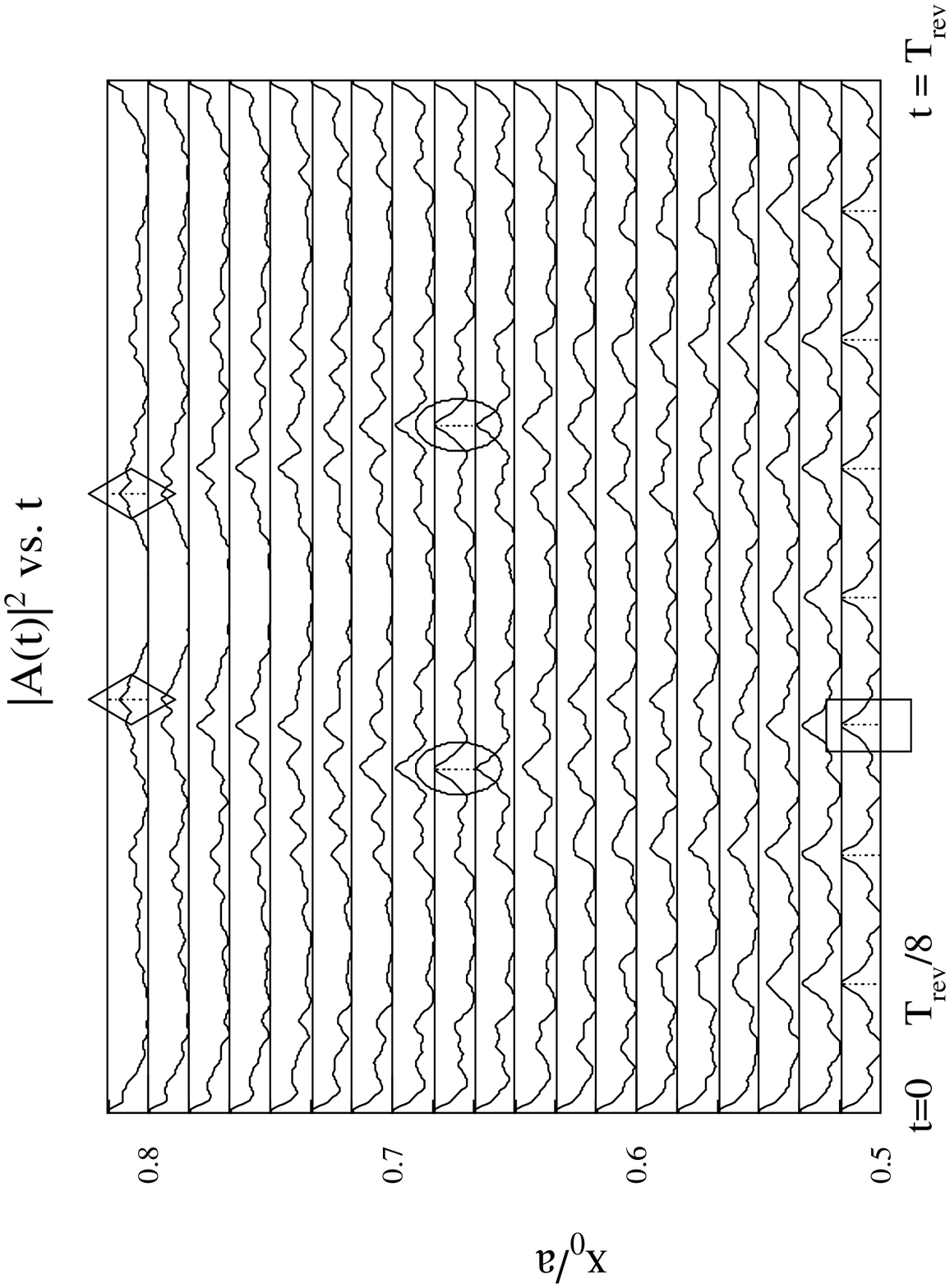,width=12cm,angle=0}
\caption{ }
\end{figure}

\newpage
\begin{figure}
\epsfig{file=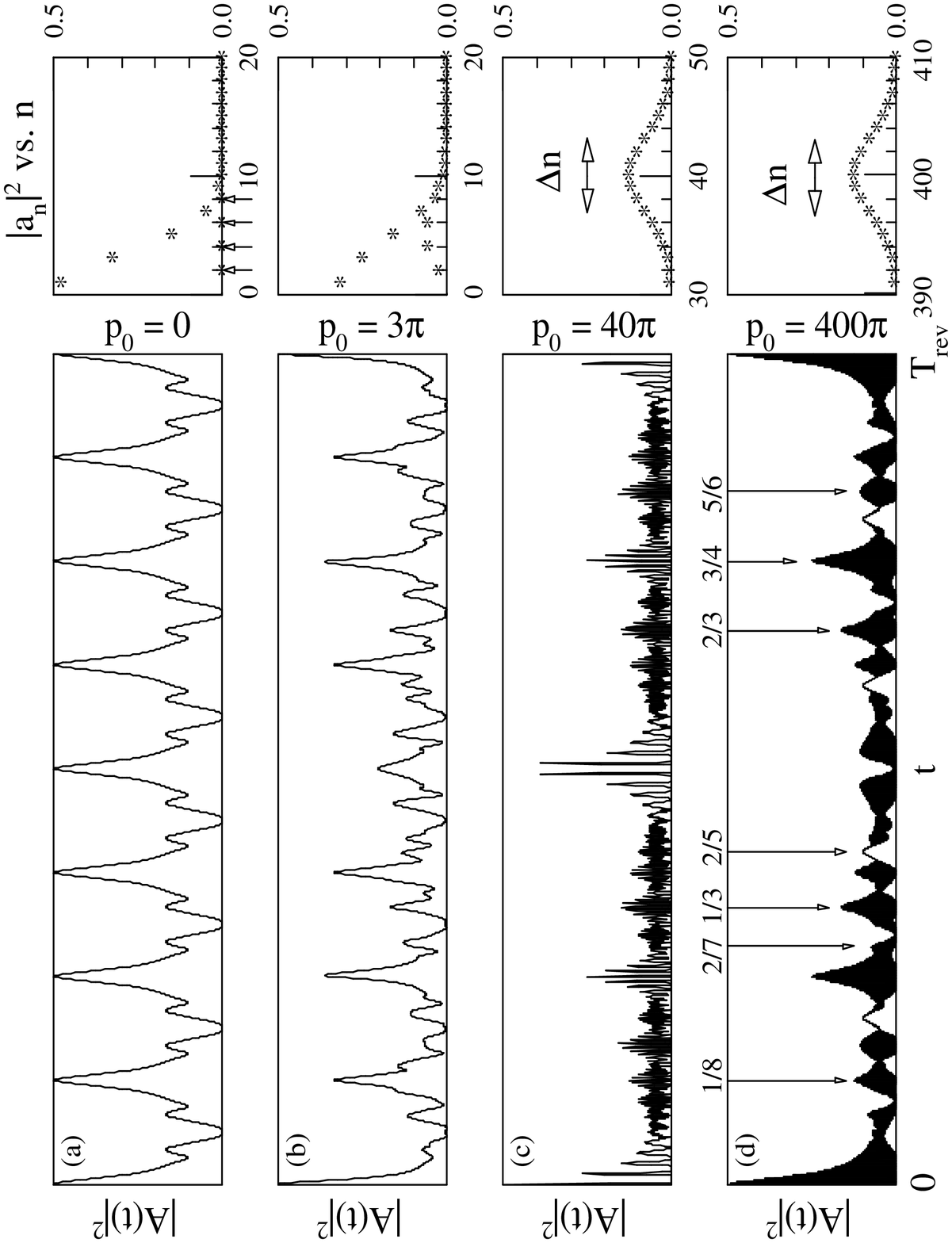,width=12cm,angle=0}
\caption{ }
\end{figure}

\newpage

\begin{figure}
\epsfig{file=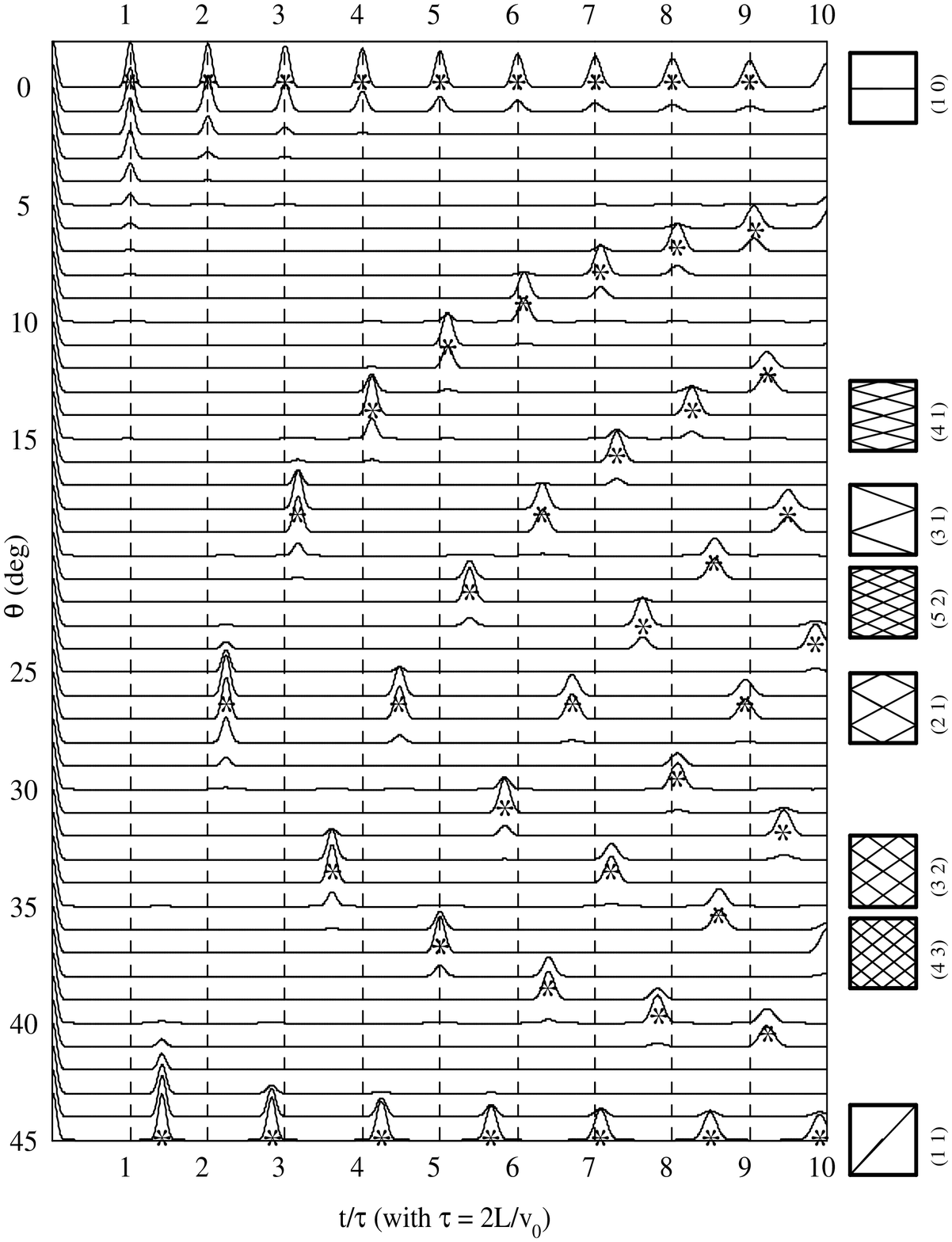,width=12cm,angle=0}
\caption{ }
\end{figure}

\newpage

\begin{figure}
\epsfig{file=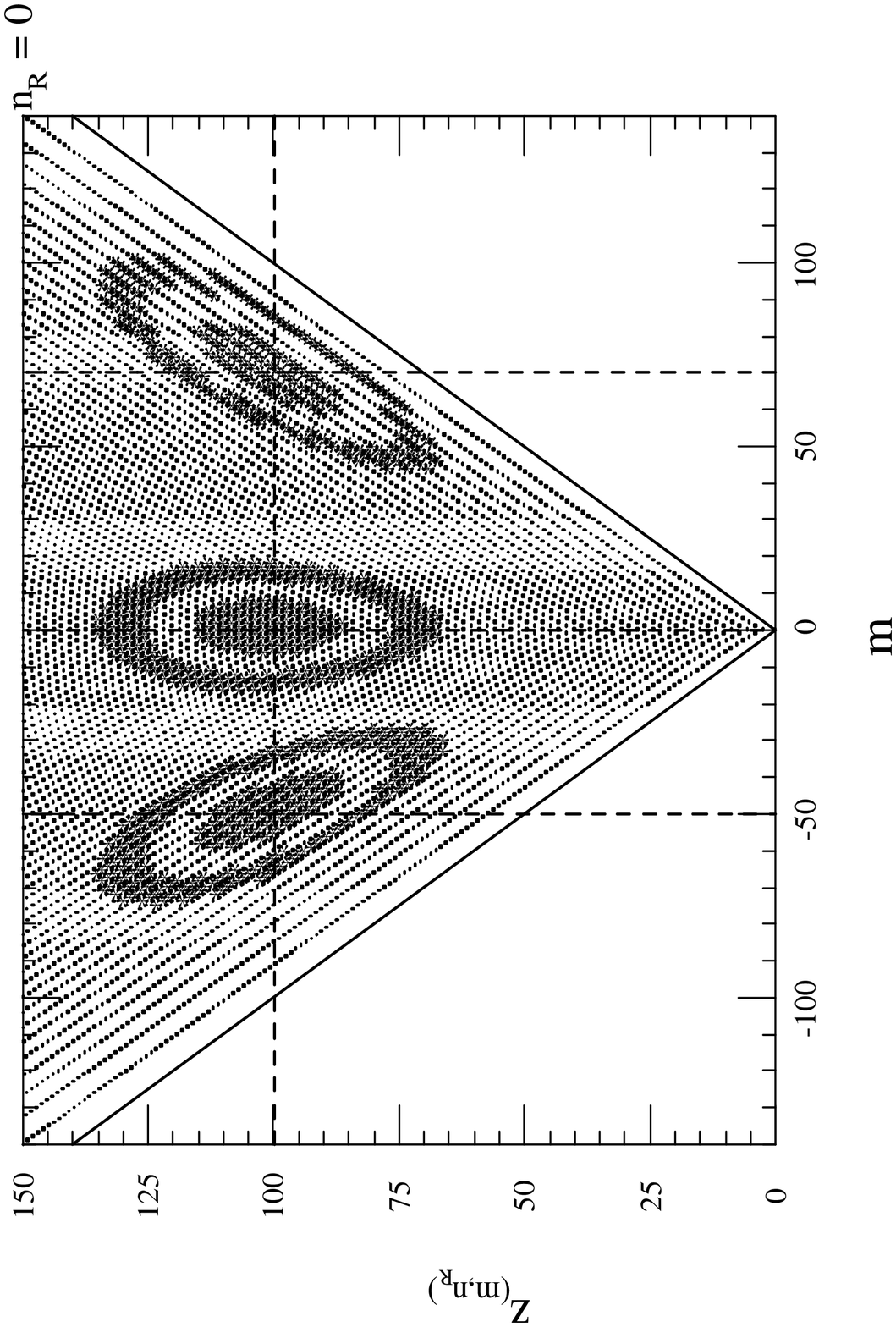,width=12cm,angle=0}
\caption{ }
\end{figure}

\newpage
\begin{figure}
\epsfig{file=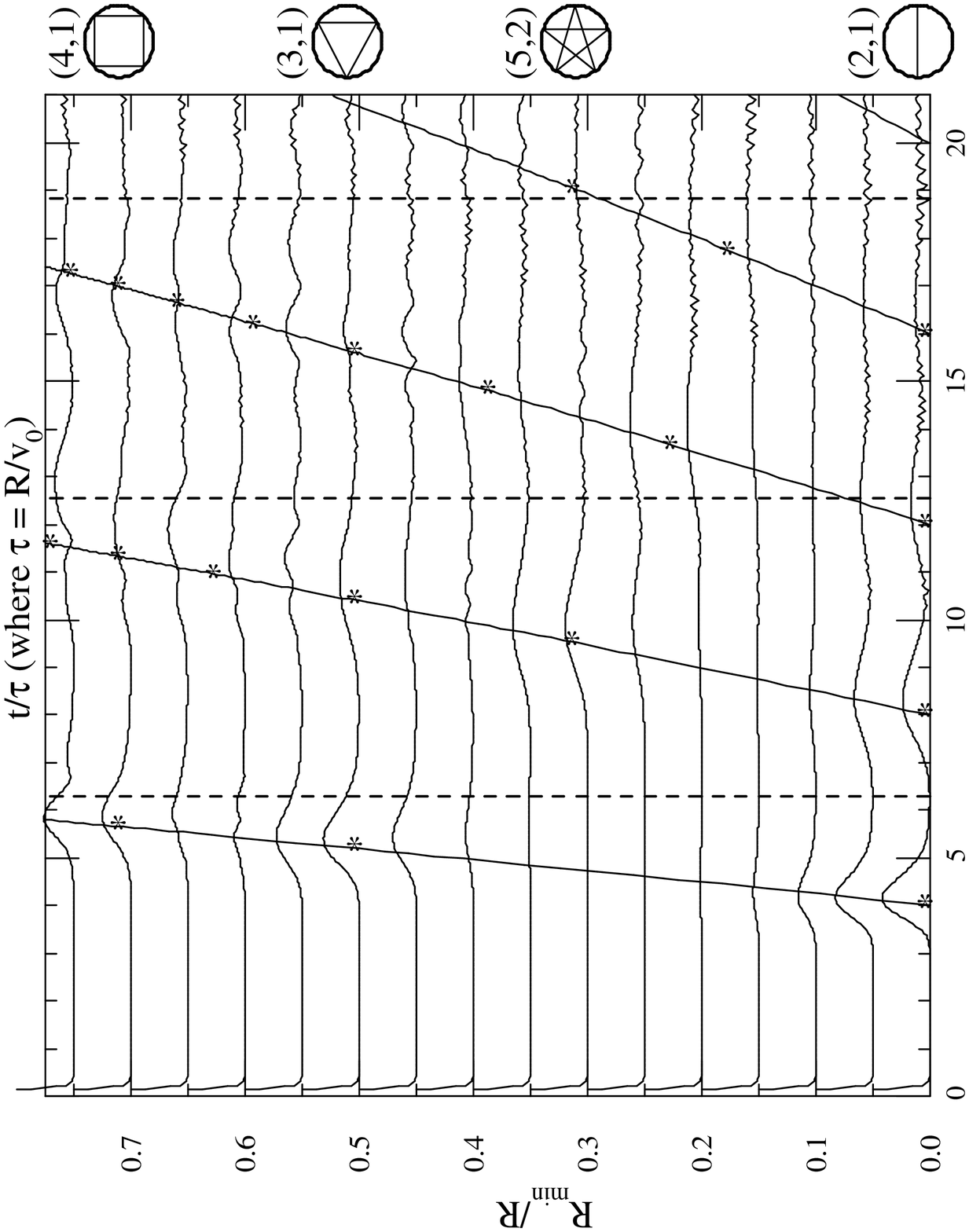,width=12cm,angle=0}
\caption{ }
\end{figure}

\end{document}